\documentclass[aps,prl,reprint,groupedaddress,showpacs,amsmath,amssymb,floatfix]{revtex4-1}
\usepackage{amsmath}
\usepackage{graphicx}
\usepackage{ulem}

\usepackage{titlesec}
\titleformat{\section}[hang]
  {\normalfont\large\bfseries}{\thesection.}{0.5em}{}
\titlespacing*{\section}{0pt}{0.75em}{0.0em}  

\renewcommand\thesubsection{S.\arabic{subsection}}
\titleformat{\subsection}
  {\normalfont\large\bfseries}{\thesubsection.}{0.5em}{}
\usepackage{bibunits} %allows for two bibliographies
\defaultbibliographystyle{apsrev4-1}

\usepackage{physics} % JACK: helpful macros, feel free to delete if conflicts with something
\newcommand{\pd}[0]{\partial}
\newcommand{\gammamr}{\gamma_{\mathrm{mr}}}
\newcommand{\gammaee}{\gamma_{\mathrm{ee}}}
\usepackage{verbatim}
\usepackage{notes2bib}

\usepackage[colorlinks=true,linkcolor=violet,citecolor=blue]{hyperref}
\usepackage{cleveref}
\usepackage[dvipsnames]{xcolor} %added by Itai

\newcommand{\jg}[1]{\textcolor{purple}{jg: #1}}
\newcommand{\andy}[1]{\textcolor{orange}{andy: #1}}

\newcommand{\Vkpfm}{V_{\textrm{KPFM}}}
\newcommand{\um}{$\mu$m }
\newcommand{\Vg}{V_{\textrm{G}}}

\newcommand{\dVdIdx}{{\partial^2 \Vkpfm / \partial x \partial I}}

\newcommand{\dVdx}{{\partial \Vkpfm / \partial x}}

\begin{document}

\title{Supersonic flow and hydraulic jump in an electronic de Laval nozzle}

\author{Johannes Geurs$^{1,2}$$^{*}$}
\author{Tatiana A. Webb$^{2,3}$$^{*}$}
\author{Yinjie Guo$^2$}
\author{Itai Keren$^2$}
\author{Jack H. Farrell$^4$}
\author{Jikai Xu$^2$}
\author{Kenji Watanabe$^5$}
\author{Takashi Taniguchi$^6$}
\author{Dmitri N. Basov$^2$}
\author{James Hone$^7$}
\author{Andrew Lucas$^4$}
\author{Abhay Pasupathy$^{2,8}$$^{\dag}$}
\author{Cory R. Dean$^2$$^{\dag}$}

\affiliation{$^{1}$Columbia Nano Initiative, Columbia University, New York}
\affiliation{$^{2}$Department of Physics, Columbia University, New York}
\affiliation{$^{3}$Department of Physics and Astronomy, Barnard College, New York}
\affiliation{$^{4}$Department of Physics and Center for Theory of Quantum Matter, University of Colorado Boulder, Colorado}
\affiliation{$^{5}$Research Center for Electronic and Optical Materials, National Institute for Materials Science, Tsukuba, Japan}
\affiliation{$^{6}$Research Center for Materials Nanoarchitectonics, National Institute for Materials Science, Tsukuba, Japan}
\affiliation{$^{7}$Department of Mechanical Engineering, Columbia University, New York}
\affiliation{$^{8}$Condensed Matter Physics and Materials Science Division, Brookhaven National Lab, New York}
\affiliation{$^{*}$Authors contributed equally to this work}
\affiliation{$^{\dag}$ Email: apn2108@columbia.edu, cd2478@columbia.edu}

\maketitle

\begin{bibunit}

%\begin{refsection} 

\textbf{
In very clean solid-state systems, where carrier-carrier interactions dominate over any other scattering mechanisms, the flow of electrons can be described within a hydrodynamic framework \cite{Gurzhi1968a, Levitov2015, lucas2018}.  In these cases, analogues of viscous fluid phenomena have been experimentally observed \cite{bandurin2016, aharon-steinberg2022a, palm2024, sulpizio2019, crossno2015, vool2021, krishnakumar2017, krebs2023, vijayakrishnan2025}.
%In very clean solid-state systems, electrons can flow like a fluid \cite{Gurzhi1968a, Levitov2015, lucas2018}.In these systems, analogues of fluid phenomena have been observed \cite{bandurin2016, aharon-steinberg2022a, palm2024, sulpizio2019, crossno2015, vool2021, krishnakumar2017, krebs2023, vijayakrishnan2025}.
However, experimental studies of electron hydrodynamics have so far been limited to the low velocity, linear response regime.
At velocities approaching the speed of sound, the electronic fluid is expected to exhibit compressible behaviour where nonlinear effects and discontinuities such as shocks and choked flow have long been predicted \cite{Dyakonov1995}.
This compressible regime remains unexplored in electronic systems \cite{hui2021a}, despite its promise of strongly nonlinear flow phenomena.
Here, we demonstrate compressible electron flow in bilayer graphene through an electronic de Laval nozzle \cite{moors2019}, a structure that accelerates charge carriers past the electronic speed of sound, until they slow down suddenly in a shock. Discontinuities in transport measurements and local flattening of potential in Kelvin probe measurements are consistent with a viscous electron shock front and the presence of supersonic electron flow, and are not consistent with Ohmic or ballistic flow. Breaking the sound barrier in electron liquids opens the door for novel, intrinsically nonlinear electronic devices beyond the paradigm of incompressible flow.
}

In solid-state physics, advances in device quality have enabled the observation of hydrodynamic flow of electrons. In very clean materials where the mean free path between the momentum-relaxing collisions of charge carriers exceeds the mean free path for momentum-conserving collisions between carriers \cite{Gurzhi1968a}, the electron system can be viewed as a collective fluid and described by a hydrodynamic model \cite{Levitov2015, lucas2018}.  In this regime, hydrodynamic models can ultimately provide general transport predictions in strongly correlated regimes that may be difficult to understand using conventional paradigms.

By now, experimental evidence for electron hydrodynamics has been found in many materials, including graphene \cite{crossno2015, bandurin2016, aharon-steinberg2022a}, GaAs \cite{vijayakrishnan2025} and  WTe$_2$ \cite{vool2021}.
Such experiments have detected whirlpools \cite{bandurin2016, aharon-steinberg2022a, palm2024} and Poiseuille flow \cite{sulpizio2019}, observed superballistic flow \cite{krishnakumar2017} and unusual thermoelectric transport \cite{crossno2015}, and measured viscous non-local transport \cite{krishnakumar2017, krebs2023} and magnetotransport \cite{berdyugin2019, zeng2024}.  All of these experiments probed subsonic flows, with small electron drift velocities $v_d$ compared to the electronic speed of sound $v_s$: the so-called incompressible flow regime where the carrier density is assumed to remain constant through the device.
A number of studies \cite{hatke2009, Andersen2019, Greenaway2021, barajas-aguilar2024, dong2025} have observed phononic effects where $v_d \sim v_{phonon}$, however the typical phonon velocities are an order of magnitude lower than the electronic sound velocity, so that these remain in the incompressible regime.
%but these did not operate in the hydrodynamic regime. Typical electronic sound velocities are an order of magnitude higher than phonon velocities.
When $v_d \sim v_s$, density fluctuations can no longer be neglected in general, and the fluid enters the compressible flow regime \cite{courant1977}. The compressible electron flow regime has never been observed \cite{hui2021a,Zhao2023}, despite predictions for novel electronic devices, such as terahertz radiation generators \cite{dyakonov1993, Dyakonov1995, mendl2021b}. The technical obstacles are: (\emph{1}) to create flows where $v_d\sim v_s$, and (\emph{2}) to find a compelling experimental signature of the compressible flow.
%\andy{static flow can be compressible at low velocity near the dirac point, so i deleted the last phrase} 

In this work, we exploit the favourable hydrodynamic properties of bilayer graphene to accelerate electrons past the speed of sound, using the de Laval nozzle geometry \cite{lifshitz1987, moors2019}. The crossover from subsonic to supersonic flow coincides with a discontinuity in the local electrochemical potential, analogous to the hydraulic jump observed in supercritical classical fluids \cite{gilmore1950}. We identify the electronic shock through combined global transport and local Kelvin probe force microscopy (KPFM), confirming the presence of compressible electron flow. The local measurements allow us to directly image the shock, appearing as a narrow front. Its position and shape are confirmed by hydrodynamic modelling.

\begin{figure*}
    \centering
    \includegraphics[width=0.85\linewidth]{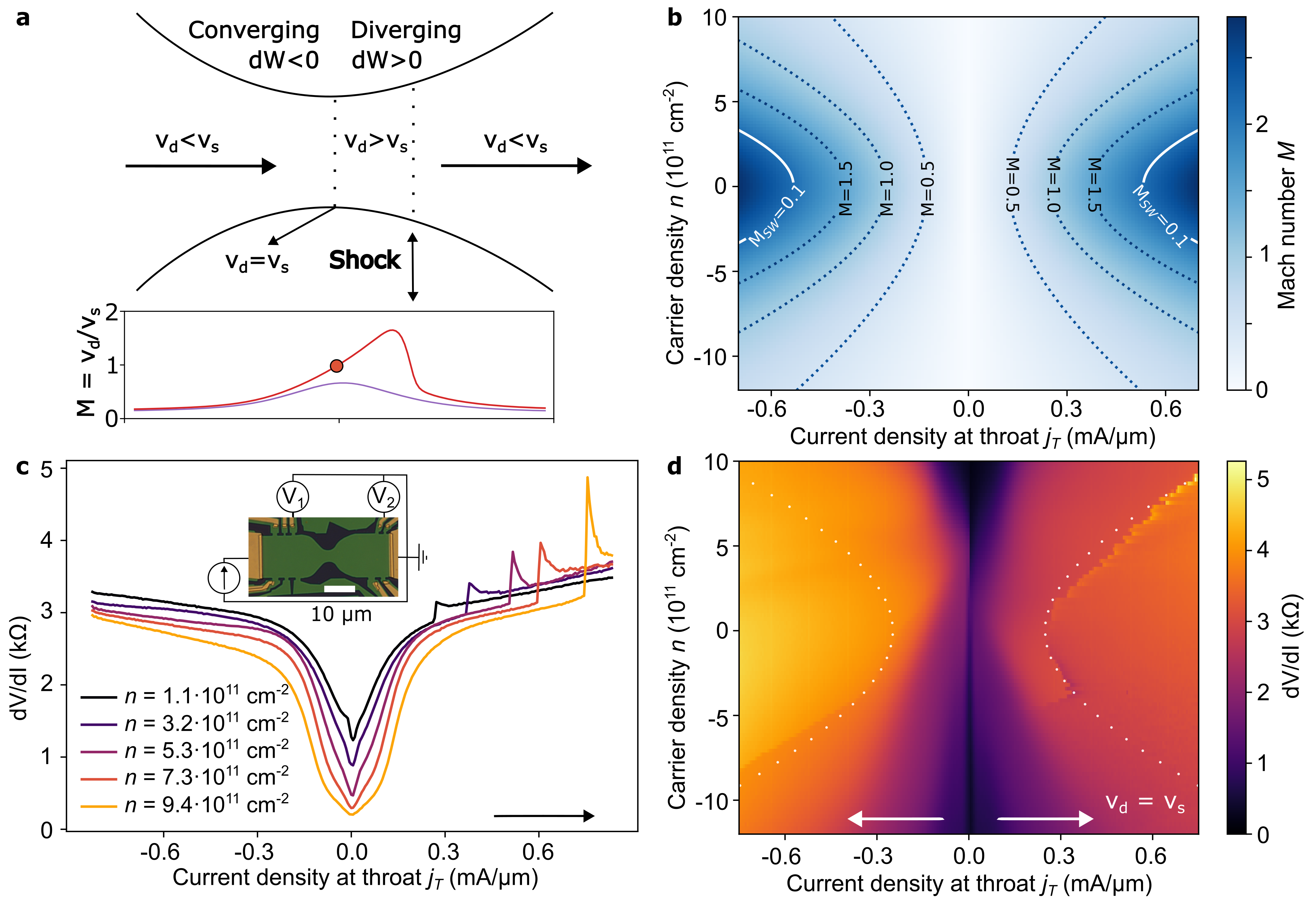}
    \caption{\textbf{An electronic de Laval nozzle.} \textbf{a.} Overview of a de Laval nozzle. Charge carriers accelerate in the converging part, reach the velocity of sound $v_s$ at the throat and then accelerate further in the diverging part, until their sudden relaxation at the shock. The graph shows the local Mach number $M=v_d/v_s$ along the constriction for the typical subsonic (purple) and supersonic (red) regimes. $M$ reaches unity at the throat in the supersonic regime (orange dot). The arrows indicate the direction of charge carrier flow.
    \textbf{b.} The Mach number $M$ that can be reached in bilayer graphene in realistic conditions (dotted lines). The solid line shows the Mach number $M_{SW}$ if the shallow water mode were present.
    \textbf{c.} Measured differential resistance of a bilayer graphene de Laval nozzle as a function of current for several gate voltages $V_G$. A sudden step in $dV/dI$ is visible on the right hand side. The arrow indicates the sweep direction. Inset: image of the device and measurement geometry.
    \textbf{d.} Measured differential resistance as a function of carrier density and current density. White dots indicate the expected location where the drift velocity reaches the electronic speed of sound at the throat (Eq. \ref{eq:sound}). Arrows indicate the sweep direction.}
    \label{fig:1}
\end{figure*}

\section*{The de Laval nozzle}
%Moors \textit{et al.} \cite{moors2019} have proposed the de Laval nozzle geometry (Fig \ref{fig:1}a) \cite{lifshitz1987} as an ideal device to demonstrate compressible electron flow by realizing an electronic shock.
Figure \ref{fig:1} shows a schematic illustration of the de Laval nozzle geometry. A current $I$ is sent through the constriction, resulting in a drift velocity $v_d = I/W(x)n(x)e$, where the shape of the constriction defines $W(x)$, $n(x)$ is the local carrier density and $e$ the elementary charge. In the incompressible limit, where $n(x)$ is constant, the local drift velocity $v_d$ is set only by the width $W$ so that it increases in the converging part and decreases in the diverging part.
In the compressible regime, $n(x)$ is allowed to fluctuate. In materials with parabolic bands, flow through a constriction is defined by the nozzle equation (derived in appendix \ref{sec:nozzlephys}):
\begin{equation}
    (M^2 - 1) \cdot \frac{dv_d}{v_d} = \frac{dW}{W},
    \label{eq:nozzle}
\end{equation}
where $M=v_d/v_s$ is the Mach number. In the subsonic regime (purple line in Fig. \ref{fig:1}a), where $M<1$, the prefactor ($M^2-1$) in Eq. \ref{eq:nozzle} is negative, such that the flow accelerates ($dv_d>0$) in the converging part ($dW<0$), and decelerates ($dv_d<0$) in the diverging part ($dW>0$), as expected from an incompressible fluid.
However, if the fluid reaches the sound velocity at the throat ($M=1$, orange dot), the prefactor changes sign, and the fluid continues to accelerate past the throat, becoming supersonic ($M>1$) in the diverging part (red line in Fig. \ref{fig:1}a). This asymmetry between compression and expansion is characteristic of compressible flow. The supersonic flow relaxes to subsonic velocities at a location \textit{past} the constriction: the shock, a stationary location where the flow parameters (velocity, carrier density) are discontinuous in the ideal hydrodynamic limit.

%In this paper, we probe compressible electron flow through an electronic de Laval nozzle. Crucially, we lower the electronic speed of sound in bilayer graphene to below the drift velocity. Global nonlinear effects at large bias are apparent in a series of discontinuities in electronic transport, consistent with the predicted electronic speed of sound. Moreover, we use Kelvin probe force microscopy (KPFM) to directly image the nonequilibrium flow, uncovering the nature of the shock as a localized nonresistive front. A hydrodynamic model of the device implies supersonic electron flow which abruptly relaxes at the shock.

The inset of Fig. \ref{fig:1}c shows an optical microscope image of the device used in this study, consisting of bilayer graphene encapsulated within $h$BN. The device must satisfy two conditions: approximate momentum conservation in the electronic fluid (hydrodynamic regime), and supersonic flow ($v_d>v_s$).
Bilayer graphene exhibits hydrodynamic behaviour over a large range of temperatures at low carrier density $n$ \cite{bandurin2016, ho2018, tan2022}, satisfying the first requirement.
To explore whether supersonic flow can be achieved, we note that the electronic sound velocity $v_s$ in an isothermal system has two components (derived in appendix \ref{sec:speedofsound}, \cite{hui2021a}): 
\begin{equation}
    v_s^2 = v_{FL}^2 + v_{SW}^2 = \frac{1}{m^*} \frac{\pd P}{\pd n} + \frac{ne^2}{m^*C},
    \label{eq:sound}
\end{equation}
where $m^*$ is the effective carrier mass, $e$ is the elementary charge, $P$ is the degeneracy pressure of the electronic Fermi gas and $C$ is the capacitance per unit area to a nearby gate. The Fermi liquid term $v_{FL}$ is the electronic analogue of the speed of sound in bulk materials. For isotropic fermionic matter this term evaluates to a constant ``cosmic sound'' $v_{FL}=v_F/\sqrt{d}$ where $v_F$ is the Fermi velocity and $d$ is the dimensionality \cite{Phan2013, Bedaque2015, lucas2016}. In bilayer graphene with quadratic dispersion, $v_{FL}(n)=\frac{\hbar \sqrt{\pi n}}{\sqrt{2}m^*}$, where $\hbar$ is the reduced Planck's constant. For a typical device with carrier density $n$ = 10$^{12}$ cm$^{-2}$,  $v_{FL}=440~$km/s. The second term $v_{SW}$ of Eq. \ref{eq:sound} arises due to the capacitive coupling $C$ between the electronic liquid and a backgate, and is the electronic analogue of waves in shallow water \cite{dyakonov1993}. In typical devices, this term is expected to provide the dominant contribution to $v_s$: for instance, a standard metallic backgate 300 nm away from an electron fluid with density $n$ = 10$^{12}$ cm$^{-2}$ leads to $v_s\approx 9500$ km/s, far beyond any typically achievable $v_d$ \cite{Yamoah2017}. Even a backgate in the atomic limit, 2 nm away, would double the speed of sound beyond the value set by $v_{FL}$ alone. This phenomenon is a consequence of the proportionality between local potential $V(x)$ and local density $n(x)$ known as the gradual channel approximation $en(x)=CV(x)$. To circumvent this obstacle, we place our devices on a lightly doped Si backgate and operate in a 10 K environment. With very long settling times, the backgate still responds to static (dc) voltages, so that the average carrier density in the channel can be controlled, but we no longer expect a local relationship between carrier density and potential, and the shallow water mode is removed from Eq.~\ref{eq:sound}.

By  eliminating the $v_{SW}$ contribution to the speed of sound, the Mach number at the throat reduces to $M=v_{d}/v_{FL}\propto I/Wn^{3/2}$.  For a fixed throat width $W_T$,  we can dynamically tune the Mach number by varying the carrier density, $n$, and drive current, $I$.  In Fig. 1b we plot a contour map of Mach numbers as a function of $n$ and current density at the throat, $j_T = I/W_T$. We restrict the current density to values that bilayer graphene are known to withstand \cite{kim2015a, moser2007}. Dashed lines mark contours of constant Mach number calculated by assuming $v_{SW}$ is eliminated and solid line is the location of $M = 0.1$ in the case of a fully metallic backgate, i.e. the shallow water mode $v_{SW}$ is present.

\begin{figure*}
    \centering
    \includegraphics[width=0.85\linewidth]{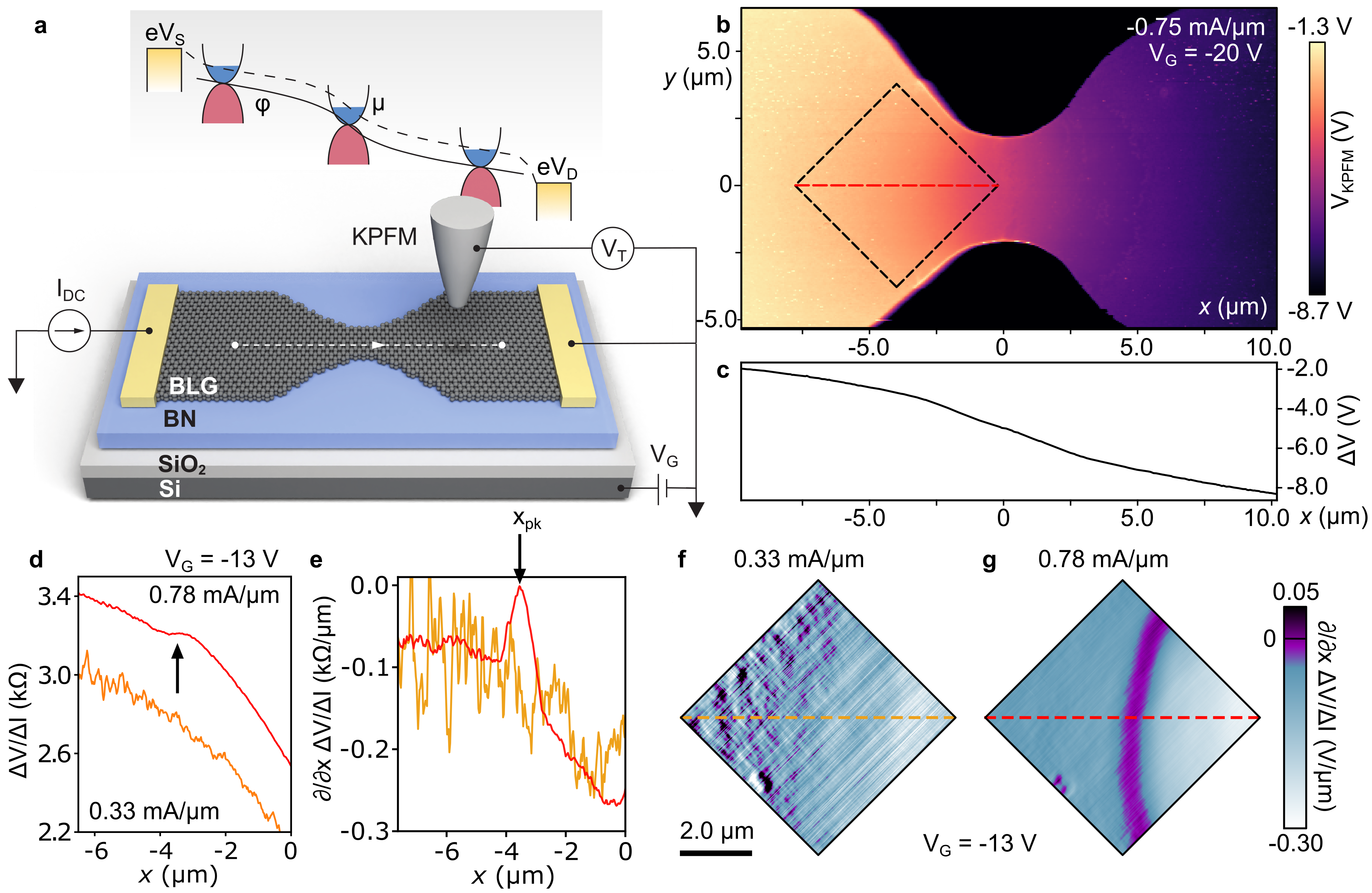}
    \caption{\textbf{Spatial profile of an electronic hydraulic jump.}
    \textbf{(a)} KPFM measurement schematic. With current flowing through the sample, biases $V_T$ and $\Vg$ are applied to the tip and gate, both referenced to the drain. The layer of BN encapsulating the bilayer graphene nozzle has been omitted for clarity.
    The influence of electric potential on the bilayer graphene bands throughout the device is shown schematically for the linear response regime.
    %\textbf{(b)} KPFM operating principle. $V_\textrm{tip}$ is adjusted to minimize the electric field between tip and sample. This value of $V_\textrm{tip}$ recorded in KPFM measurements is denoted as $\Vkpfm$. Because of the current in the sample, the Fermi level $E_F(X)$ changes spatially throughout the sample. $W_\textrm{tip}$ and $W(x)$ are the tip and sample work functions, respectively.
    \textbf{(b)} KPFM image with $\Vg=~$-20~V, $j=~$-0.75~mA/$\mu$m. The values outside of the bilayer graphene nozzle exceed the colour scale limits as a result of the gate. %{Source Q7, drain Q3}. 
    \textbf{(c)} Local potential $\Delta V$ along the device in panel b.
    \textbf{(d)} Local potential $\Delta V$ along the constriction (red dashed line in b) for two current densities. $\Delta V = \Vkpfm(I) - \Vkpfm(0.28 \textrm{mA}/\mu m)$ removes work function features, and $\Delta I$ is the change in current from the $j$ = 0.28 mA/\um reference measurement. Near the critical current density (red), a flattening of the potential is visible.
    \textbf{(e)} The spatial derivative $\frac{d}{dx}\frac{\Delta V}{\Delta I}$, calculated from the same data set as (d). A clear maximum to zero at $x_{pk}$ marks the flattening of the potential.
    \textbf{(f,g)} $\frac{d}{dx}\frac{\Delta V}{\Delta I}$ approximately in the square area marked in (c). Refer to methods for details of obtaining  $\frac{d}{dx}\frac{\Delta V}{\Delta I}$ from KPFM images. }
    %\textbf{(e)} Profile of $\Delta V / \Delta I$ along the lines indicated in (c, d), where $\Delta I= I - 1.0\textrm{mA}$.
    \label{fig:2}
\end{figure*}

\begin{figure*}
    \centering
    \includegraphics[width=0.8\textwidth]{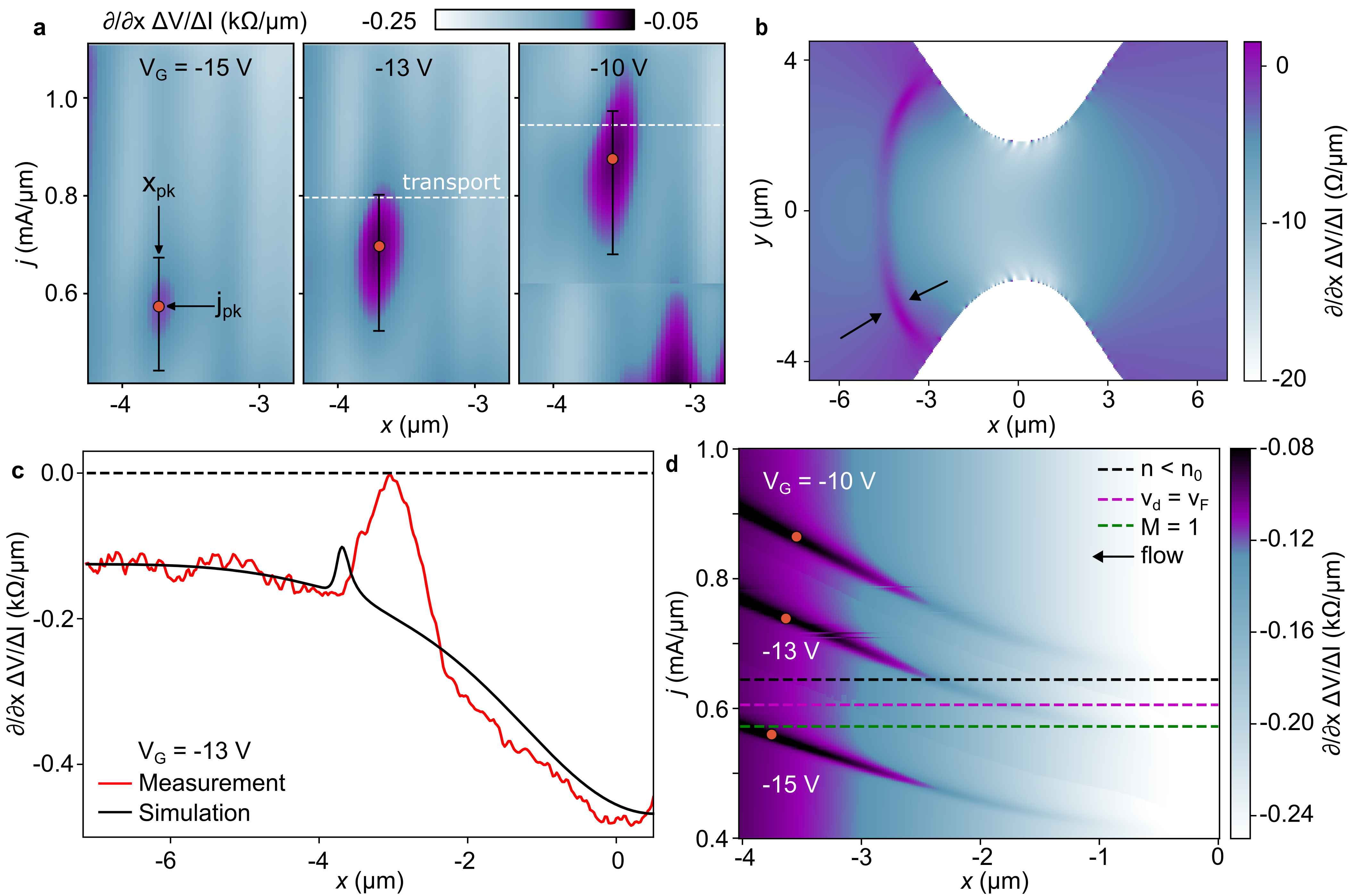}
    \caption{\textbf{Extent of an electronic shock.}
    \textbf{(a)} Measured local differential resistance $\frac{\partial}{\partial x} \frac{\Delta V}{\Delta I}(x,j)$ shows that the hydraulic jump (maximum in dark purple) is present only for a small range of current densities and positions. The red dashed line indicates the current density at which in situ transport shows maximal $dV/dI$.
    \textbf{(b)} The calculated local differential resistance $\frac{\partial}{\partial x} \frac{\Delta V}{\Delta I}$ for the nozzle in the supersonic regime, matching the experimental conditions at $V_G=-13V$. The shock is visible as the arc downstream from the throat, matching the measured profile of Fig.~\ref{fig:2}g.
    \textbf{(c)} Comparison of the numerical solution to Eq.~\ref{eq:1dns}, to the measured $\frac{d}{dx} \frac{\Delta V}{\Delta I}(x,j)$ of Fig. \ref{fig:2}e. At current density $j$ = 0.78 mA/$\mu$m, the calculated location of the shock is close to the observed maximum.
    \textbf{(d)} Composite image of numerical solutions for the shock at three different gate voltages. The shocks, visible as the maxima in $\frac{\partial}{\partial x} \frac{\Delta V}{\Delta I}(x,j)$, become stronger and travel downstream from the throat (at $x$ = 0) as applied current density increases. The green dashed line indicates the sonic point in the $V_G=-13V$ case, the lowest $j$ at which $M = 1$ is reached at the throat. The other two dashed lines show possible limits to supersonic flow for the $V_G=-13V$ case: situations where $v_d$ reaches $v_F$ downstream (purple) or $n$ is depleted below the thermal density $n_0$ (black).
    }
    \label{fig:3}
\end{figure*}
\section*{Transport discontinuities}
To probe global electron flow, the differential resistance $dV/dI$ of the nozzle as a function of current density at the throat $j_T$ is measured at several gate voltages. The voltage difference is measured between two contacts on either side of the constriction in a four-probe configuration (inset of Fig. \ref{fig:1}c). At small $j_T$, the profile of $dV/dI$ increases quadratically with $j_T$ (Fig. \ref{fig:1}c). This is a consequence of Joule heating, which was previously measured in graphene Hall bars \cite{huang2023}. We estimate (appendix \ref{locdiscontinuity}) that the electron temperature reaches 150 K at high bias, bringing the device in the hydrodynamic regime \cite{ho2018}.

At higher $j_T$, a discontinuity appears as a sudden increase in differential resistance. The discontinuity is only present in one current sweep direction (arrow) and moves to higher $j_T$ as the carrier density $n$ is increased.
The map of $dV/dI$ in Fig. \ref{fig:1}d, measured while sweeping away from zero bias (arrows), shows the discontinuity for both electrons and holes.
The white dashed line indicates where the drift velocity is expected to reach the electronic velocity of sound in our device, according to Eq. \ref{eq:sound} (appendix \ref{locdiscontinuity}). The close match between the discontinuity in $dV/dI$ and the prediction for $M=1$ suggests a phenomenon connected to $v_s = v_{FL}$, and indicates that the shallow water mode has successfully been avoided.
The discontinuity is only partially present for hole type carriers in forwards bias, and is absent for electron type carriers in reverse bias. 
Another device (appendix \ref{transporttests}) show discontinuities for both carrier types, but only for reverse bias. These variations in bias dependence might be attributed to a variation in device quality between devices or homogeneity between left and right sides of a single device.
We note that this transport feature is absent, under any bias condition, in fixed-width bilayer graphene Hall bars with no constriction, confirming the role of the de Laval nozzle geometry (appendix \ref{transporttests}).   Additionally, transport measurements through a constriction in monolayer graphene also did not display any evidence of a discontinuity (appendix \ref{transporttests}). This is expected since in monolayer graphene the minimal sound velocity is 700 km/s. 

\section*{Hydraulic jump}
The discontinuous transition from supercritical to subcritical flow in classical thin films manifests as a hydraulic jump \cite{bhagat2018}.  A familiar example is observed when water flowing from a faucet hits a flat surface and the resulting thin film spreads out faster than the shallow water wave velocity. The sudden transition from supercritical, shooting flow to subcritical, slow flow is directly visible as a height difference. The height difference results from the sudden conversion of kinetic energy to gravitational potential energy.
%The discontinuous transition between supersonic and subsonic flow in a de Laval nozzle is analogous to the well-known hydraulic jump \cite{bhagat2018}. This phenomenon, occurring when flow from a faucet hits a flat surface, demonstrates the sudden transition from supercritical shooting flow to slow, subcritical flow. The critical velocity in this case is the velocity of shallow water waves. A discontinuity in potential energy accompanies the discontinuity in velocity, directly visible as a height difference. 
In our system, we expect a discontinuity in potential energy for the electronic shock too, visible as a discontinuity in the local electrochemical potential in the region where the flow expands, which we can measure with Kelvin probe force microscopy (KPFM).The analogy between shocks and hydraulic jumps is only exact for fluids with heat capacity ratio $\gamma = 2$ \cite{gilmore1950}. By the equipartition theorem, this is the case for electrons with  parabolic band structure in two dimensions.

%In the absence of current, KPFM measures the local contact potential difference between sample and tip: $\Vkpfm=W(x) - W_\textrm{tip}$, where $W(x)$ and $W_\textrm{tip}$ are the sample and tip work functions (Figure \ref{fig:3}b).
Figure~\ref{fig:2}a depicts the measurement setup. KPFM is sensitive to both electric potential and chemical potential (Fermi energy) which vary spatially when a current is flowing \cite{yu2009}, as illustrated in the inset in panel a. Figure~\ref{fig:2}b is an example spatial $\Vkpfm$ map of the device under bias, in a regime where no transport discontinuity is observed. The signal is dominated by an Ohmic electrochemical potential profile, giving a nearly 8~V drop across the field of view, with the steepest slope at the constriction (panel c).
We can separate electronic effects from smaller local work function variations due to impurities on the surface (appendix \ref{SI:work}) by considering the difference in $\Vkpfm$ at different $I$, denoted by $\Delta V$.

\section*{Local measurements}
Figure \ref{fig:2}d shows line cuts along the dashed line in Fig. \ref{fig:2}b below (yellow) and near (red) current density where simultaneous transport measurements observe a maximum in $dV/dI$. The yellow profile has a monotonic dependence on position, as expected from Ohmic flow. The red curve has a pronounced flattening of the potential, indicated by the arrow. The calculated spatial derivative $\frac{d}{dx}\frac{\Delta V}{\Delta I}$ (Fig. \ref{fig:2}e) shows a vanishingly small local differential resistance (arrow at $x_{pk}$). When this quantity is mapped over the device (rhombus in Fig. \ref{fig:2}f,g corresponds to the location in panel c), an arc shaped band of flattened local potential is present at a current density near that of the anomalous transport features of Fig. \ref{fig:1}c,d. We claim that this is an electronic hydraulic jump, caused by supersonic electron flow between the throat and $x_{pk}$.

Additional measurements (appendix \ref{simul}) confirm that this flattening is the source of the transport discontinuity and support its interpretation as a hydraulic jump at a point \textit{past} the constriction (towards the source for electron-type carriers). The hydraulic jump moves to the other side of the throat when source and drain are swapped, and disappears when a perpendicular magnetic field is applied, as expected from a hydrodynamic phenomenon.

Fig. \ref{fig:3}a maps the extent of the hydraulic jump by plotting the local differential resistance $\frac{d}{dx}\frac{\Delta V}{\Delta I}$ as a function of position $x$ and current density $j$, measured along the centre line of the device (red dashed line in Fig. \ref{fig:2}b). The three panels show the evolution of the hydraulic jump at three different carrier densities, as set by the gate voltage. The extent of the hydraulic jump, which appears as a local maximum in $\frac{d}{dx}\frac{\Delta V}{\Delta I}$ at a narrow range of current densities and positions, is indicated by the bars. As gate voltage decreases, the region of the hydraulic jump moves to higher $j$, with the upper bound closely matching the location of the peak in the global $dV/dI$ measurements (white dashed line, appendix \ref{insitu}).

\section*{Modelling compressible flow}
To understand the behaviour of electronic shocks, we perform extensive numerical simulations of compressible electron flow with a minimal hydrodynamic model (appendix \ref{numerical}). The relevant degrees of freedom are the electron density $n(\vb x)$, drift velocity $\vb{v}_d(\vb x)$ and electrostatic potential $\phi(\vb x)$. %for the Fermi liquid in a constriction at high bias, which applies for regions of $(V_g, I_{SD})$ parameter space above or below charge neutrality, so that one species of charge carrier realizes---we take positive carriers for simplicity. \footnote{across CNP, maintaining the sign of current $I_{SD}$ requires large temperature gradients, obscuring the shock physics and invalidating the isothermal $T\approx 0$ modelling} In that case, the Fermi Liquid phase, an appropriate set of fields with which to model the experiment consists the local number density of charge carriers $n(\vb x, t)$, the local carrier drift velocity $\vb {v}_d$, and the local electric potential $\phi(\vb x)$, presumed quasistatic. % The electric current density is $\vb{J}_e\equiv \pm e J$, depending on the sign of the charge carriers.
The equation of motion for carrier density $n$ is simply the continuity equation expressing conservation of electric charge.  The equation of motion for $\vb{v}_d$ represents a balance of several fluxes and forces: drag from collisions at rate $\gammamr (n)$, electrostatic force $\grad \phi(\vb x)$; viscous drag from carrier-carrier interactions with viscosity $\eta$; gradients in the local electronic pressure $P(n) = \hbar^2 \pi n^2 / 4 m$ (appendix~(\ref{sec:speedofsound})); and, finally, nonlinear convection.  The results are the Navier-Stokes Equations:
\begin{subequations}\label{eq:ns}
\begin{gather}
    \pd_t n + \div \vb (n \vb{v_d}) = 0, \label{eq:continuity} \\
    n\, \qty(\pd_t \vb v_d + \vb{v}_d \cdot \grad \vb{v}_d) + \frac{1}{m}\qty(\grad P + \eta \grad^2 \vb{v_d} - e\, n\, \grad\phi) \notag\\
    = -\gammamr (n)n \vb{v}_d.
\end{gather}
\end{subequations}
 %\andy{I don't like this wording -- the previous experiments use models where $P$ (due to $n$) is slowly varying and this can account for the observed data.  I think we should instead say something like: $\phi$ accounts for the fact that the voltage associated to the CNP is drifting throughout the device, which must be significant to explain the data?}\jf{not convinced CNP is moving. I don't think the presence of an external driving field is too mysterious. in 1D transport, we always add a constant E. in the nozzle, constant E is not right because not consistent with Laplace, the driving external field looks more like $E \sim 1 / w$.} \andy{to say another way, in the past when we were always thinking about linear response, it didn't really matter your interpretation because you only coupled to $\nabla (\delta \mu - \delta \phi)$, the overall electrochemical potential gradient.  But here we are sensitive to which part is $\delta \mu$ and which part is $\delta \phi$ because of the nonlinearity in $n$.  }\jf{that's true. But then again at reasonable carrier densities (though maybe a bit higher than we had hoped) the magnitude of the hump in KPFM seems about right, though there's a slight mismatch\ldots}

The resulting spatially resolved local differential resistance $\frac{\partial}{\partial x}\frac{\Delta V}{\Delta I}$ is shown in Fig. \ref{fig:3}b, for parameters $j, n$ matching the experiment in Fig. \ref{fig:2}g. The shock is visible as an arc (arrows). Its location matches the observations well, and the spatial width of the hydraulic jump is set by a length scale $l_{ee} \sim$ 100 nm \cite{lifshitz1987}, related to the electronic viscosity by $\eta = m^*n v_F \ell_{\mathrm{ee}}/4$ \cite{lucas2018}. From a typical $\dv*{V}{I}$ near $I = 0$ of $300\ \Omega$ (Fig.~\ref{fig:1}b), we estimate a corresponding momentum relaxing mean free path $\ell_{\mathrm{mr}} \sim 2\ \mu\mathrm{m}$, comparable to the throat width, placing the device squarely in the quasihydrodynamic regime. We used this low value of $\gammamr$ in the two-dimensional simulation.  However, the large measured differential resistance at large bias suggests a much faster rate of momentum relaxation, $\ell_{\mathrm mr} \sim 200$ nm. For compressible flow to persist, this additional resistance cannot correspond to a constant $\gammamr$. We suggest that a nonlinear $\gamma(n) = \alpha n$, for constant $\alpha$, allows high resistance and compressible flow while additionally accounting for the weak dependence of $\dv*{V}{I}$ on density ($V_G$) in Fig.~\ref{fig:1}c. Assuming quasi one-dimensional flow \cite{courant1977}, we solved Eqs.~\eqref{eq:ns} with $\gammamr(n) = \alpha n$ (appendix \ref{sec:1dmodel}) showing agreement with the measured resistance and shock location (Fig.~\ref{fig:3}c). 
It is worth noting that the experiment gets tantalizingly close to a sign reversal of $\frac{\partial}{\partial x}\frac{\Delta V}{\Delta I}$ (dashed line in Fig. \ref{fig:3}c is zero). This is allowed for strong shocks \cite{moors2019} but would require higher device quality (appendix \ref{numerical}) or higher current densities.

In Fig.~\ref{fig:3}d we plot three solutions of Eqs.~\ref{eq:ns} as a function of position and current density, corresponding to the three gate voltages of the measurements in panel a.
For low current bias the shocks are weak and located near the throat.  With increasing bias the shock magnitude grows and shift further downstream with the shock persisting for arbitrarily large current.
The experiment in panel a does not capture the movement of the shocks, but instead shows shocks appearing at a single distance from the throat, for a narrow range of currents. While thermal smearing likely obscures the weak shocks close to the nozzle throat, and sample disorder could pin the location of the shock to a point in the device, the sudden disappearance of the shock in Fig. \ref{fig:3}a, combined with an increased $dV/dI$ in transport, points to a mechanism that is not present in the unipolar model.

We discuss two scenarios as possible endpoints of an electronic shock. (i) As the area with supersonic flow increases, the local density $n(x)$ downstream must decrease to accommodate increasing drift velocity. The local carrier density cannot be depleted below the thermal density $n_0$; when $n$ reaches this value, the temperature is comparable to Fermi temperature and our model is no longer applicable. %\textcolor{red}{jh: how does the physics change if the model is no longer applicable?  i.e. does the velocity fall below the unipolar limit?  does electron-hole scattering become dominant and increase resistance?}. \jg{I wish we knew! Conductivity around charge neutrality also involves the incoherent conductivity.}
Reaching this neutrality point could also explain why the end of the shock is marked by an increase in $dV/dI$ in transport (white dashed line in Fig.~\ref{fig:3}a): the device is minimally conductive in the areas where $n(x) \approx n_0$.  (ii) The Fermi velocity sets a limit to the velocity of a charge carrier. Previous studies indicate a new dissipative mechanism in this case \cite{Berdyugin2022, nowakowski2025a} and report a similar increase in $dV/dI$. An additional limit to supersonic flow could be the length scale over which momentum is relaxed within the fluid, $l_{\mathrm{mr}}$.
The dashed lines in Fig.~\ref{fig:3}d show the lower and upper bounds of an electronic shock for the $V_G=-13V$ case, although the experimental location and current density do not match (orange dots). A full description of compressible electron flow, away from the unipolar limit, would be required.

In conclusion, we have realized the de Laval nozzle geometry in an electronic system. The electron fluid is accelerated past its speed of sound when flowing through a constriction, until its sudden deceleration at a shock. This electronic hydraulic jump is visible in locally resolved KPFM measurements and its endpoint coincides with an excess resistance observed in transport. The realization of compressible electron hydrodynamics in a two-dimensional electronic system could enable a number of practical possibilities, such as terahertz radiation generation \cite{dyakonov1993, farrell2022}. It also allows mapping known inherently nonlinear, compressible phenomena in classical fluids onto electronic devices with completely new working principles \cite{hui2021a}.
An analogy by Unruh \cite{unruh1981} treats the de Laval nozzle as an \textit{acoustic} black hole \cite{novello2002, volovik2009, braunstein2023}. For electronic fluids, the large speed of sound would lead to a relatively large Hawking temperature $T_H \sim$ 500 mK.
Unusual transport phenomena may be realized in the compressible flow regime in quantum materials that break time-reversal or rotational symmetries \cite{cook2019}, have relativistic dispersions \cite{crossno2015} or in chiral fluids, which can arise in models of nuclear physics \cite{bernhard2019, son2009}.

%\newpage
\section*{Acknowledgements}
The authors acknowledge helpful discussions with Michael Hilke, Michael Fogler, Leonid Levitov and Vijay Vedula.

CRD and JH were supported by Gordon and Betty Moore Foundation’s EPiQS Initiative, grant GBMF1027.  J.H.F. and A.L. were supported by the Gordon and Betty Moore Foundation's EPiQS Initiative under Grant GBMF10279 and the National Science Foundation under CAREER Award DMR-2145544. DNB is an EPIQS investigator in quantum materials GBMF945.  Additional sample fabrication (J.G., J.H.) was supported by the ARO MURI program (W911NF-21-2-0147).  KPFM measurements were supported by the Air Force Office of Scientific Research via award FA9550-21-1-0378. Scan probe data analysis was supported by the Scan Probe Microscopy Group at Brookhaven National Laboratory under Department of Energy Contract No. DE-SC0012704. Synthesis of hBN (K.W., T.T.) was supported by the JSPS KAKENHI (Grant Numbers 21H05233 and 23H02052), the CREST (JPMJCR24A5), JST and World Premier International Research Center Initiative (WPI), MEXT, Japan.
The development of the Columbia multimodal nano imaging platform is supported  as part of Programmable Quantum Materials, an Energy Frontier Research Center funded by the U.S. Department of Energy (DOE), Office of Science, Basic Energy Sciences (BES), under award DE-SC0019443. The authors acknowledge the use of facilities and instrumentation supported by the NSF through the Columbia MRSEC (DMR-2011738) and computing resources from Columbia University's Shared Research Computing Facility.

\section*{Author contributions}
J.G. conceived the project. Y.G. and J.G. fabricated the device. J.G. performed electronic transport measurements. T.W. and I.K. performed Kelvin probe measurements. J.H.F. and A.L. provided theory and numerical simulations. K.W. and T.T. grew hBN crystals. J.G. wrote the manuscript with input from all authors.

%\newpage
\section*{Methods}

%\subsection*{Sample fabrication}
\textbf{Sample fabrication:}
We prepare all graphene and hBN flakes on SiO$_2$/Si chips by exfoliating from bulk crystals. The layer number of graphene is identified by the colour contrast from the optical microscope. We use a thin polycarbonate film to pick up the exfoliated layers with the dry transfer method \cite{wang2013}. The stack is picked up in the sequence of top hBN/graphene/bottom hBN and dropped onto a 285nm SiO$_2$/Si substrate at 180 $^{\circ}$C. A lightly doped Si substrate ($\rho$ = 3.5-6.5 $\Omega$cm) was chosen so that the dopants in the Si freeze out at cryogenic temperatures.

Metal contacts to the graphene are deposited after etching through the top BN layer with SF$_6$. Finally the device is etched into the proper shape (nozzle) by alternating SF$_6$ and O$_2$ plasma. The top surface of the device is cleaned by atomic force microscope (Bruker) in contact mode \cite{lindvall2012a}.

%\subsection*{Transport setup}
\medskip
\noindent
\textbf{Transport measurement:}
Transport measurements were performed in a cryogenic system (Oxford Teslatron) in the standard four probe configuration. The devices were biased with a voltage-controlled current source (Stanford Research CS580) and read out by a digital multimeter (Agilent 34401A). A DC gate voltage was applied by a digital source (Keithley 2400 SMU).

%\subsection*{Kelvin probe force microscopy measurements}
\medskip
\noindent
\textbf{Kelvin probe force microscopy}
Sideband Kelvin probe force microscopy (KPFM) measurements were performed in an Attocube cantilever-based cryogenic atomic force microscope with interferometric detection (attoAFM I with attoLIQUID 2000 cryostat), operating in a helium exchange gas. Measurements used Nanosensors PPP-EFM probes with resonant frequency $f_\circ$ near 61~kHz and coated with a double layer of chromium and platinum iridium.

Sideband KPFM measurements were performed by applying an AC bias modulation on the probe (frequency $f_{\textrm{AC}} = 800$~Hz and amplitude in the range 0.5~V to 2.0~V). The cantilever deflection signal was demodulated at frequency $f_\circ + f_{\textrm{AC}}$ using a Zurich Instruments HF2LI lockin amplifier. A control loop was then used to adjust the probe DC bias to minimize the cantilever response at $f_\circ + f_{\textrm{AC}}$. The probe bias was then recorded while scanning the tip across the sample surface in amplitude-controlled topographic feedback, with setpoint ranging from 50~nm to 120~nm.

The probe DC bias is limited to $\pm 10 V_{DC}$ as a system limitation. However, at high sample biases, the local electrochemical potential regularly exceeds this limit. In these cases, the drain $V_D$ was offset by a static DC voltage. This shifts all other potentials in the system so that the local potential falls within the probe bias limit. In the main text we adopt the convention that device or gate potentials $V_x$ refer to $V_x-V_D$ even in the case where a constant offset $V_D$ was necessary.

The $\frac{\partial}{\partial x} \Delta V / \Delta I$ images in Fig.~\ref{fig:2}f-g were produced as follows. A series of KPFM ($\Vkpfm(I, x, y)$) images was recorded over the same area for several different $I$ values. The DC current through the device, $I$ was held fixed during each image. The series of images was drift corrected by first scaling then shifting to align a set of features manually chosen from the simultaneously recorded topographic images. The scaling accounts for piezo scanner calibration changes due to the microscope temperature changing with the applied current. $\Delta V(I, x, y) = \Vkpfm(I, x, y) - \Vkpfm(1.0\textrm{mA}, x, y)$ was calculated to remove work function features, and then the numerical derivative was calculated along the x direction, followed by smoothing with a uniform filter of width 0.5~$\mu$m to reduce the appearance of noise. The profiles shown in Fig.~\ref{fig:2}d-e are taken from these images before (d) and after (e) differentiation. Normalization by $\Delta I = I - 1.0$~mA facilitates comparison across $I$.

To examine the $I$-dependence of the potential, $\Vkpfm(I, x)$ data sets were recorded by continuously scanning the probe back and forth along the $x$ axis of the nozzle, while slowly ramping $I$. $\Vkpfm(I, x)$ was smoothed using a Gaussian filter with $\sigma_x=0.12$~$\mu$m and $\sigma_I=0.08$~mA (unless otherwise noted) to reduce noise. $\Delta V$ was calculated by subtracting the $I=0$ scan line. Then the numerical derivative along the $x$ direction was calculated to generate $\partial \Delta V / \partial x$, followed by dividing each scan line by the average $I$ for the line to give $\partial \Delta V / \partial x / \Delta I$ shown in Fig.~\ref{fig:3}. In this case, $\Delta I = I$ because the reference data is at $I=0$. $\dVdIdx$ was calculated by taking the numerical derivatives of the smoothed smoothed $\Vkpfm(I, x)$, and the $I$-direction differentiation conveniently reduces work function features.

For these $\Vkpfm(I, x)$ data sets, the zero of the x axis was placed at the maximum of {$\dVdIdx(x)$} for large $I$, which we expect to approximately locate the narrowest constriction of the nozzle.

We note that while we have used differences in $\Vkpfm$ measured at different $I$ to remove work function contributions to the data, this removal is imperfect due to drift, exacerbated by temperature variations at high currents, and especially in the $\Vkpfm(I, x)$ data sets where drift correction is not feasible.

\putbib[nozzle-bibtex]
\end{bibunit}

%\bibliography{nozzle,jack}

\newpage

\newpage
\clearpage

%%%% Supplementary Information
\pagebreak
\begin{widetext}
\renewcommand{\theequation}{S\arabic{equation}}
\setcounter{equation}{0}
\begin{bibunit}

\section*{Supplementary Materials}

\subsection{Review of nozzle physics} \label{sec:nozzlephys}
In this section, we review the operating principles of a de Laval nozzle~\cite{moors2019} in the context of the quasihydrodynamic toy model of \eqref{eq:ns}.  For theoretical insight, we may approximately analyze \eqref{eq:ns} by eliminating the $y$-dependence of all fields, that is, taking as an Ansatz
\begin{subequations}
    \begin{align}
        n &= n(x,t),\\
        \vb{v}_{d} &= v_d(x,t)\, \vu{x}.
    \end{align}
\end{subequations}
The one-dimensional approximation is valid in the limit of a long, thin channel, where the $y$-dependence of the fields is weak, and away from the boundaries where no-slip boundary conditions demand a narrow boundary layer of width proportional to $\eta$~\cite{lifshitz1987}. Supposing that the channel has variable width $W(x)$, conservation of quasiparticle number demands the continuity equation
\begin{align}\label{eq:1dcont}
    \pd_t(nW) + \pd_x(n v_d W) = 0,
\end{align}
and force-balance gives the Navier-Stokes equation:
\begin{gather}\label{eq:1dns}
\pd_t(nv_dW) + \pd_x \qty(n v_d^2 W) + \frac{W}{m}\pd_x P(n) -  \frac{\eta}{m} \pd_x (W \pd_x v_d) = -\gammamr nv_d W.
\end{gather}
%\begin{gather}
%    n\pd_t v_d + n v_d \pd_x v_d + \frac{1}{m} \pd_x P(n) + \frac{\eta}{m} \pd_x^2 v_d = -\gammamr nv_d.
%\end{gather}
The one-dimensional model of \eqref{eq:1dcont}, \eqref{eq:1dns} was used to create the theory plots of panels cd in Fig.~\ref{fig:3}.  We describe the numerical method in Sec.~\ref{numerical}; for now, however, we may press on analytically.  At this level, we may solve for a steady state solution with $\pd_t n = \pd_t v_d = 0$. First, recall that $v_s^2 = 1/m\pdv*{P}{n}$, meaning that $1/m \pd_x(P) = v_s^2 \pd_x n$.  Second, notice that the steady state solution to the continuity equation gives
\begin{align}
    n = \frac{I_p}{v_d W},
\end{align} 
where $I_{p}$, essentially a constant of integration, is the particle flux through the device, related to electrical current by $I_p = I/(\pm e)$. As a further simplification, let $M(x) \equiv v_d / v_s$ be the local Mach number of the fluid.  The two Navier-Stokes equations become an ordinary differential equation for $v_d$, namely
\begin{align}
\left(M^2 - 1\right)\frac{\pd_x v_d}{v_d} - \frac{\pd_x W}{W} =  \left(\frac{I_p}{W v_d}\right) \qty(\frac{\eta}{m} \pd_x ( W \pd_x v_d) - \gammamr v_d).
\end{align}
Taking $\eta = \gammamr = 0$ sets the right hand side to zero, yielding the simple ordinary differential equation Eq. \ref{eq:nozzle} of the main text, namely:
\begin{align}\label{eq:simplenozzle}
    \qty(M^2 - 1) \frac{\pd_x v_d}{v_d} = \frac{\pd_xW}{W}.
\end{align}

When $M = 1$, that is $v_d = v_s$, this differential equation is singular, and solutions on either side of such a `sonic' point must be be matched using appropriate boundary conditions.  

This nozzle equation \eqref{eq:simplenozzle} differs from that of equation 3 in Ref. \cite{moors2019}, since those authors studied monolayer graphene which features a linear (relativistic) dispersion.

\clearpage

\subsection{Speed of sound in bilayer graphene}
In this section we calculate the speed of sound $v_{FL}$ in bilayer graphene in the Fermi liquid regime with carrier density $n$ and temperature $T$. The speed of sound in a medium characterizes the velocity of compression or expansion waves in the medium.  For example, in the solid state context, acoustic phonons represent sound waves of the crystal lattice.  In this paper, we are concerned instead with the speed of sound of the Fermi liquid itself, that is, the characteristic velocity of fluctuations in the quasiparticle real-space density function.   

In bilayer graphene, bands are parabolic at low energies~\cite{Li2016a}, meaning the dispersion relation for n-type carriers is
\begin{align}\label{eq:dispersion}
    \varepsilon(\mathbf k) = \frac{\hbar^2 k^2}{2m^*}.
\end{align}
Accordingly, the density of states $\nu$ is a constant given by
\begin{align}
    \nu = \frac{gm^*}{2 \pi \hbar^2},
\end{align}
where $g \equiv g_s g_v$ is the total spin-valley degeneracy, with $g = 4$ appropriate for $A-B$-stacked bilayer graphene.

The speed of sound of the Fermi liquid $v_{FL}$ is related to its pressure $P$ by the thermodynamic relation
\begin{align}
    v_{FL}^2 = \left.\frac{1}{m^*}\pdv{P}{n}\right|_{\frac{s}{n}},
\end{align}
where the derivative is calculated at fixed entropy per particle~\cite{lucas2018}.  In the grand canonical ensemble, relevant thermodynamic quantities (density $n$, pressure $P$, energy density $\mathcal E$, entropy density $s$, are given by standard formulae (specific to $d = 2$ and parabolic band):
\begin{subequations}
\begin{align}
    n &= g\int \frac{\dd[2]{\vb k}}{(2 \pi)^2} \frac{1}{\mathrm e^{\beta(\varepsilon(\vb k) - \mu)} + 1} = \nu (k_B T)\log(1 + \mathrm{e}^{\beta \mu}), \label{eq:nmu} \\
    P &= g(k_B T) \int \frac{\dd[2]{\vb k}}{(2\pi)^2} \log(1 + \mathrm{e}^{\beta(\mu - \varepsilon(\vb k)}) = -\nu (k_B T)^2\, \mathrm{Li}_2\qty(-\mathrm{e}^{\beta \mu}),\label{eq:fullP}\\
    \mathcal E &= g\int \frac{\dd[2]{\vb k}}{(2\pi)^2} \frac{\varepsilon(\vb k)}{\mathrm e^{\beta(\varepsilon(\vb k) - \mu)} + 1} = P, \\
    s &= \frac{\mathcal E + P - \mu n}{T} = \frac{2P - \mu n}{T}.
\end{align}
\end{subequations}
With $\mathrm{Li}_2$ known as the dilogarithm function. Note that in the pressure $P$ we have dropped an additive constant. 

We notice now that in this theory, owing to the constant density of states $\nu$, entropy per particle $s / n$ depends only on the quantity $\beta \mu$.  Thus curves of constant $s/n$ correspond to curves of constant $\beta \mu$, and we may calculate the speed of sound straightforwardly (with the given choice of additive constant in the pressure) as
\begin{align}
    v_{FL}^2 = \frac{1}{m^*} \left.\pdv{P}{n}\right|_{\beta \mu} = \frac{\left.\pdv{P}{T}\right|_{\beta \mu}}{\left.\pdv{n}{T}\right|_{\beta\mu}} = \frac{2P}{m^* n}.
\end{align}
Inverting \eqref{eq:nmu} to find $\mu(n,T)$, the resulting speed of sound in the bilayer-graphene Fermi Liquid $v^2_{FL}$ is
\begin{align}\label{eq:fullvs}
    v_{FL}^2 = -2\frac{\nu (k_B T)^2}{m^*} \frac{\mathrm{Li}_2\qty(1-\mathrm{e}^{n / (\nu k_B T)})}{n}.
\end{align}
\label{sec:speedofsound}
\newpage
\subsubsection{Low temperature limit}
For simplicity in numerical solution to the Navier-Stokes equations in the device, we take the $T = 0$ limit of \eqref{eq:fullP} and \eqref{eq:fullvs}. The results, appropriate when $T \ll T_F$ where $T_F$ is the Fermi temperature, are
\begin{subequations}
\begin{align}
    P(n, T = 0) = \frac{n^2}{2 \nu}, \\
    v_{FL}^2(n, T = 0) = \frac{n}{m^*\nu}.
\end{align}
\end{subequations}
The quadratic pressure as a function of carrier density offers an exact analogy to the shallow-water equations of fluid dynamics, and we exploited this relationship in the numerical solutions.

\subsubsection{Back-gate modification to speed of sound}

The capacitive gate with capacitance $C$ per unit area leads to an additional electrostatic energy density $\mathcal{E}_g$ given by:
\begin{align}\label{eq:Eg}
    \mathcal{E}_g = \frac{(ne)^2}{2C}.
\end{align}
This energy density offers an additional ``shallow water'' speed of sound $v_{SW}$ given by
\begin{align}
v_{SW}^2 =\frac{e^2 n}{mC},
\end{align}
which is Eq. \ref{eq:sound} of the main text. Crucially, the speed of sound is minimized at low carrier densities $n$, making supersonic flow realistic in bilayer graphene when devices are operated at low carrier density. In the present experiment, $v_{SW}$ was eliminated by operating on a lightly doped silicon backgate near $T=0$.

\clearpage

\subsection{Effect of thermal density on the speed of sound}
\label{locdiscontinuity}
\begin{figure}
    \centering
    \includegraphics[width=0.9\linewidth]{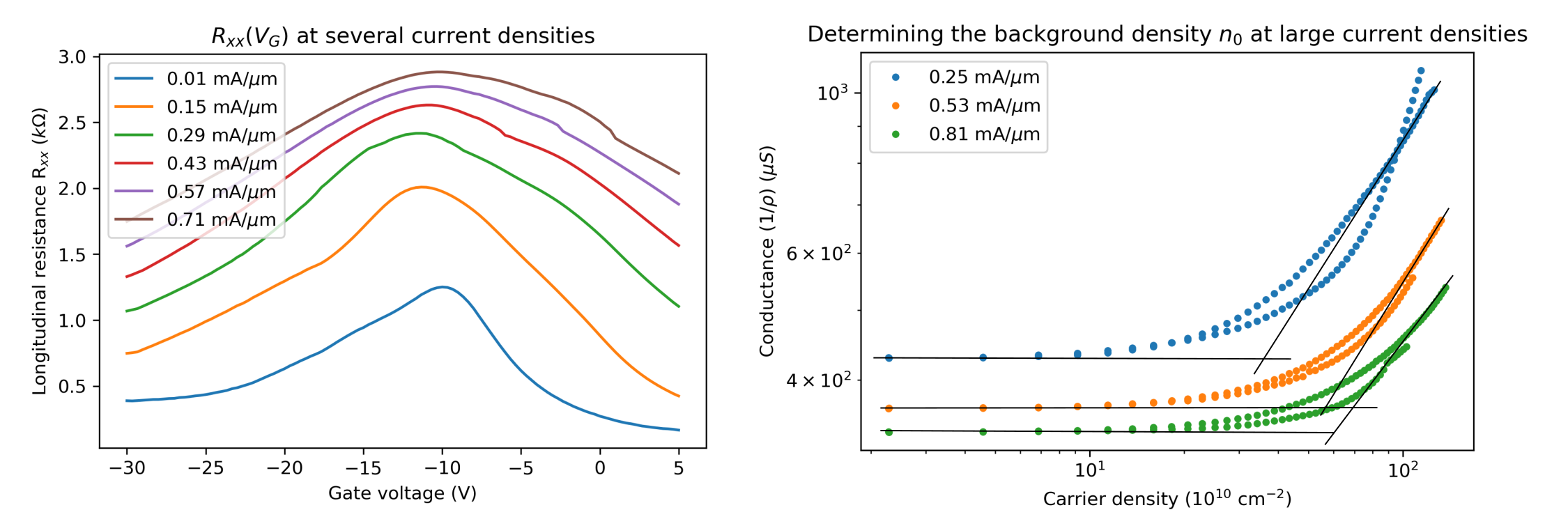}
    \caption{\textbf{Background density $n_0$.} \textbf{a.} Resistance $R$ as a function of gate voltage, at several current densities. \textbf{b.} The same data on a logarithmic scale indicates a value for $n_0$.}
    \label{fig:S1}
\end{figure}
The measured $R(V_G)$ at several current densities of the bilayer graphene device is shown in Fig. \ref{fig:S1}. The charge neutrality peak is broadened as the current density increases, likely due to thermal effects. This background density $n_0$ is the minimal carrier density present in the device. The same data, plotted as a conductance $S = 1/R$ versus density $|n|$, allows us to extract a value $n_0 \approx 5\cdot 10^{11} cm^{-2}$ at large current density.

This background density, a mixture of electron and hole carriers, must be included in the expression for the electronic speed of sound (Eq. \ref{eq:sound}) by the usual quadrature \cite{Meric2008} $n^2=n_0^2+\left(C\cdot (V_G-V_0)\right)^2$. It is visible in \ref{fig:2}b as the minimal speed of sound $v_{s,min}=\frac{\hbar \sqrt{\pi n_0}}{\sqrt{2}}$. In Fig. \ref{fig:2} the static gate offset was obtained as the gate voltage at which R($\Vg$) is maximal at 0.25 mA/$\mu$m.
%\jg{perhaps this manual insertion of $n_0$ should happen in the derivation of $v_s$}\jf{I wouldn't say so, since we're supposing there is still a sharp FS, so the equation holds with $n = \text{total carrier density}$.  If we think the temperature is high enough that these thermal carriers are behaving like an ideal gas, then indeed we'd have to modify the equation for vs, it would be $v_s^2 = k_b T / m + \hbar^2 \pi n / 2 m$}  

For completeness, we note that the standard phenomenological quadrature rule incorporating background density may be understood as a simple approximation to \eqref{eq:fullvs}, calibrated for the Fermi Liquid regime (high $n$, or low $T$). To lowest nontrivial order in $T$, \eqref{eq:fullvs} is
\begin{align}
    v^2_{FL} \approx \frac{1}{m^* \nu} \qty(n + \frac{\pi^2}{3}\frac{(\nu k_B T)^2}{n}) \approx \frac{1}{m^*\nu} \sqrt{n^2 + \frac{2 \pi^2}{3} (\nu k_B T)^2},
\end{align}
where the rightmost equality holds to the same order in $T$.  Comparing to the quadrature rule, we find $n_0 = \sqrt{2 \pi^2/3}\, (\nu k_B T)$ setting the background density scale that should be used in the quadrature approximation.  On the other hand, the thermal density (at $\mu=0$) from \eqref{eq:nmu} is $n_T = \nu k_B T \log 2$. The prefactor $\log 2$ differs from $\sqrt{2 \pi^2 / 3}$ by a factor $\approx 3.7$.  Experimentally, the measured $n_0$, added in quadrature, matches the data near $\mu = 0$---see Fig.~\ref{fig:1}d.

The measured thermal density $n_0$ allows us to estimate the electron temperature to be on the order of 150 K at high applied bias currents. This is expected from Joule heating \cite{huang2023} and would bring bilayer graphene in the hydrodynamic regime \cite{ho2018, tan2022}.

\clearpage

\subsection{Additional transport tests}
\label{transporttests}
\begin{figure}
    \centering
    \includegraphics[width=1.0\linewidth]{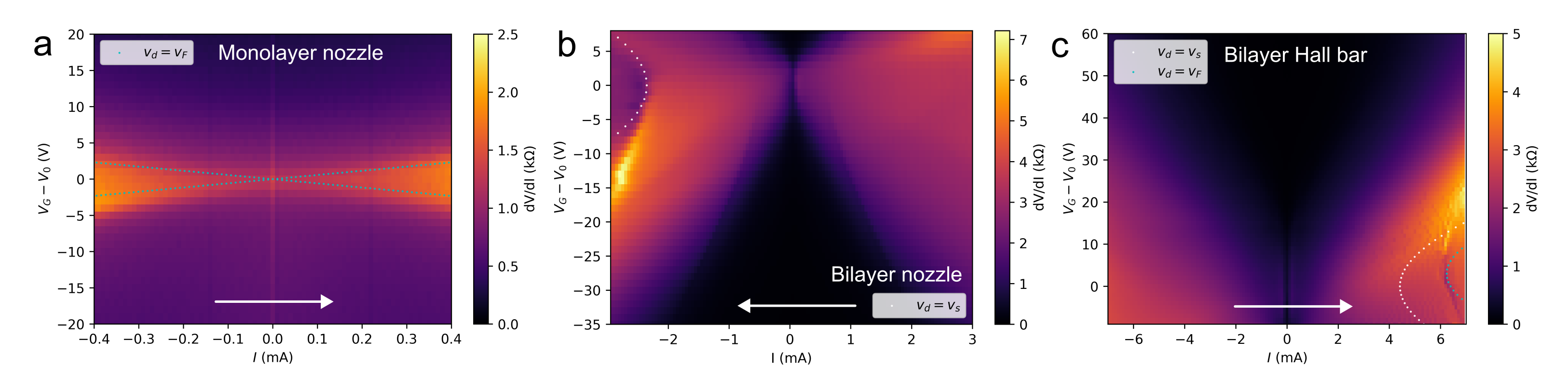}
    \caption{\textbf{Additional transport tests.}
    \textbf{(a)} Map of $dV/dI (I,V_G)$ for a monolayer graphene de Laval nozzle. The increase in $dV/dI$ follows $v_d=v_F$ (black dots), as expected from the Schwinger effect in monolayer graphene.
    \textbf{(b)} Map of $dV/dI (I,V_G)$ for an additional bilayer graphene de Laval nozzle. The discontinuity in $dV/dI$ is visible on the left side, at $v_d = v_s$.
    \textbf{(c)} Map of $dV/dI (I,V_G)$ for a bilayer graphene Hall bar. The discontinuity in $dV/dI$ is present only at $v_d=v_F$, not at $v_d = v_F$.
    The current sweep direction in all plots is indicated by the white arrow.}
    \label{fig:S2}
\end{figure}

Fig.~\ref{fig:S2}a shows $dV/dI$ measurements for a monolayer graphene nozzle. A series of $dV/dI$ maxima are present, occurring at $v_d=v_F$ (green dots in Fig. \ref{fig:S2}c) \cite{Berdyugin2022}.

Fig. \ref{fig:S2}b and c show a $dV/dI$ measurement in a bilayer graphene de Laval nozzle and Hall bar, respectively. The two devices were lithographically defined out of a single piece of bilayer graphene to isolate the effect of the geometry of the device on the measurements. The sudden change in $dV/dI$ occurs at $v_d = v_s$ for the nozzle, but only at $v_d = v_F$ for the Hall bar. This confirms the role of the geometry, and suggests that the physics at $v_F$ in the Hall bar is \textit{accelerated} in the nozzle structure.

\begin{figure}
    \centering
    \includegraphics[width=0.95\linewidth]{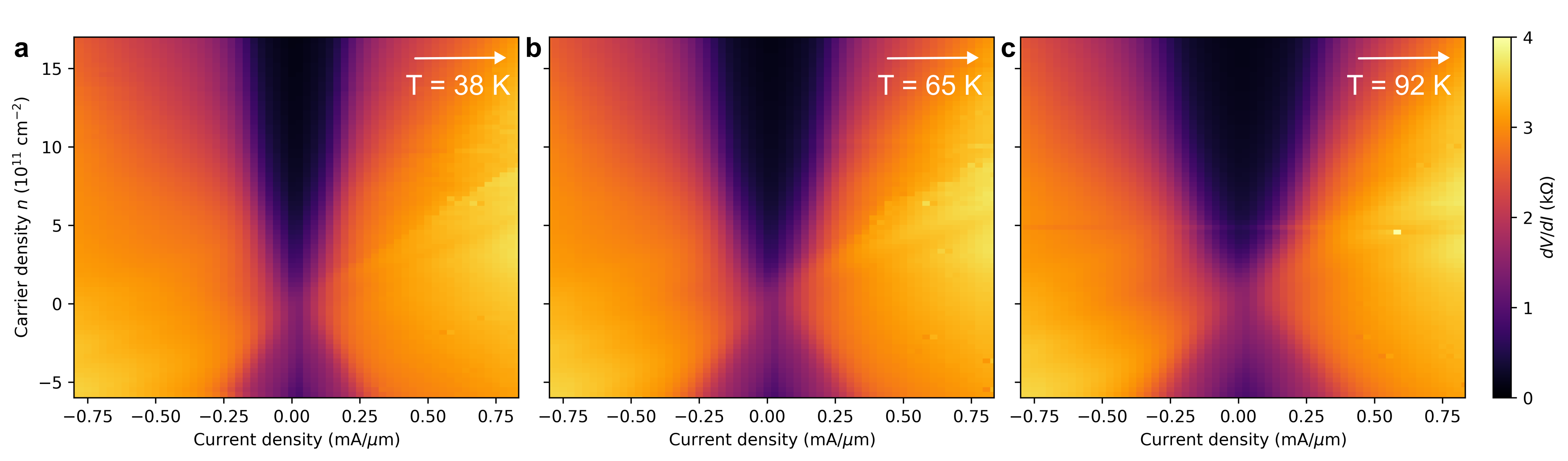}
    \caption{
    \textbf{Measurements at elevated temperature.}
    Measurements of $dV/dI (j,n)$ at elevated temperature show that the features in Fig.~\ref{fig:1}d gradually disappear. The current sweep direction is indicated by the arrow.
    }
    \label{fig:S7}
\end{figure}

Fig.~\ref{fig:S7} shows the $dV/dI (j,n)$ measurement of Fig.~\ref{fig:1}d at multiple temperatures. The discontinuity weakens and eventually disappears by $T$ = 92 K. We attribute this to the backgate becoming more conductive as the temperature is increased, resulting in the shallow water mode increasing the velocity of sound (Eq.~\ref{eq:sound}).

\clearpage

\subsection{In-situ transport measurements}
\label{insitu}
\begin{figure}[h!]
    \centering
    \label{fig:S5}
    \includegraphics[width=0.9\linewidth]{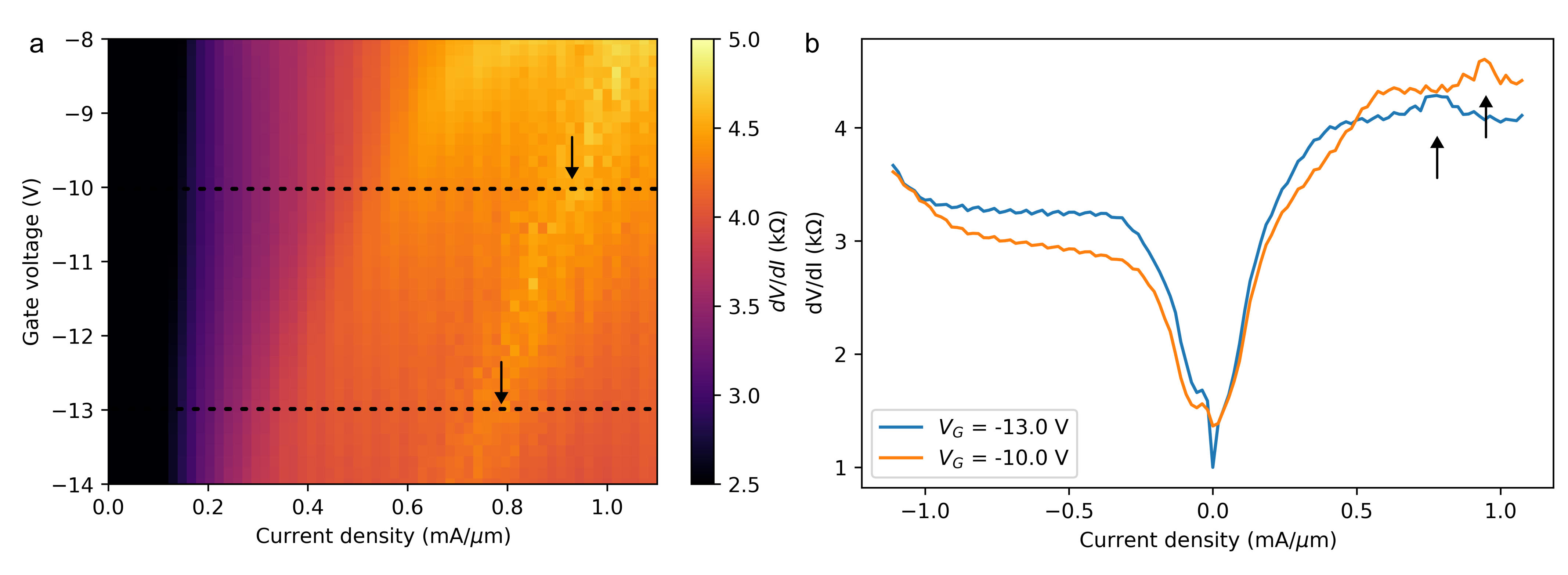}
    \caption{\textbf{In situ transport.} \textbf{a.} Map of $dV/dI (I,V_G$. Black dashed line indicates where two cuts are taken. \textbf{b.} Cuts of $dV/dI$ at constant gate voltage. The location of the peak in $dV/dI$ is indicated by an arrow.}
\end{figure}

Transport quantities, such as the voltage drop along the sample and derived quantities such as $dV/dI$, can still be measured during KPFM measurements. The series of peaks in $dV/dI$ visible in Fig. \ref{fig:S5}a have been measured just before the KPFM measurements of Fig. \ref{fig:2}. In particular, the current density at the $dV/dI$ maximum (black arrows in Fig. \ref{fig:S5}) is indicated by the white dashed lines in Fig. \ref{fig:3}a.
\clearpage

\subsection{Impact of carrier density and B on shock} \label{morekpfm}
\begin{figure}[h!]
    \centering
    \label{fig:Scompare}
    \includegraphics[width=0.9\linewidth]{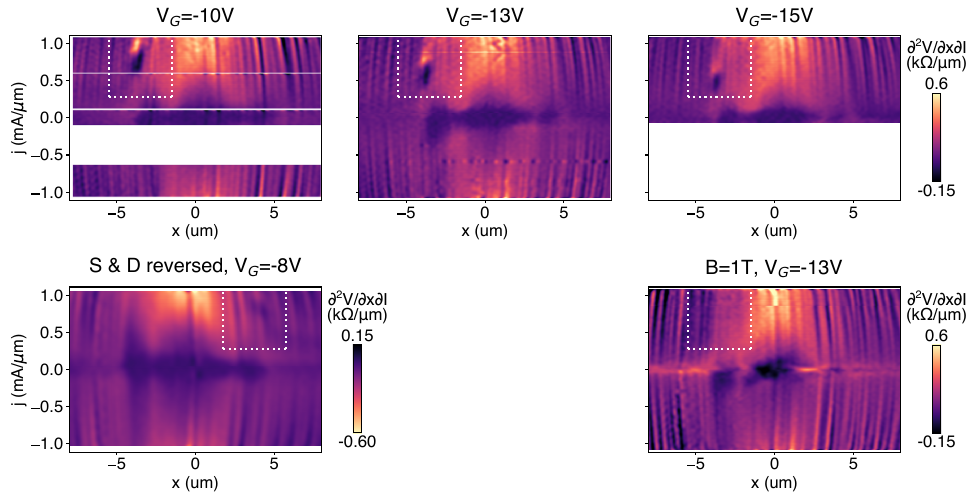}
    \caption{\textbf{Additional KPFM results.} The top row are $\dVdIdx$ maps at different gate voltages along the centre of the nozzle. On the bottom left, the $\dVdIdx$ map has source and drain reversed (smoothed with $\sigma_x$ =0.2~$\mu$m and $\sigma_j$ = 0.04~mA/$\mu $m). The bottom right map was measured with an applied perpendicular magnetic field of 1 T, where the Lorentz force breaks momentum conservation. The typical dip-hump feature of the hydraulic jump is visible in the white dashed boxes in all cases, except when $B$ = 1T.}
\end{figure}

\begin{figure}[h!]
    \centering
    \label{fig:Szoom}
    \includegraphics[width=0.9\linewidth]{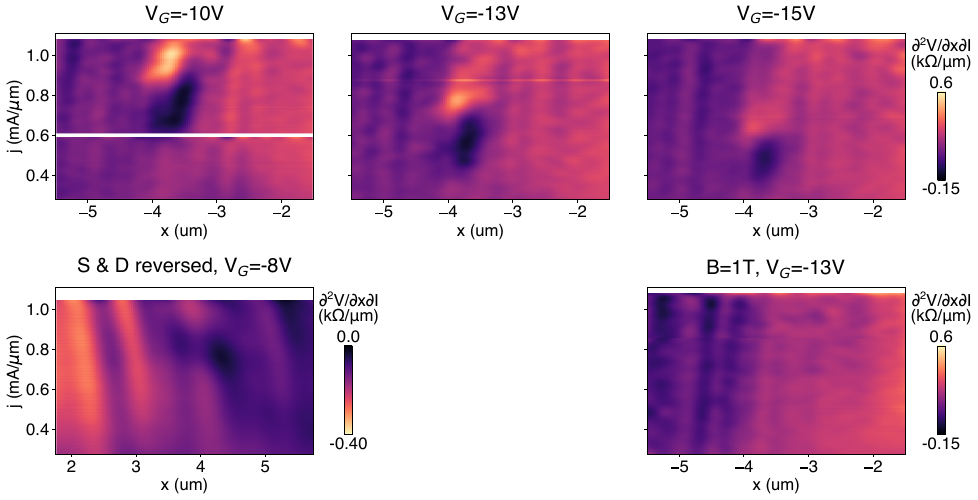}
    \caption{$\dVdIdx$ maps, magnifications of the areas in the dashed boxes of Fig. \ref{fig:Scompare}.}
\end{figure}

The hydraulic jump is very clearly visible when the $\frac{\partial}{\partial x}\frac{\Delta V}{\Delta I}$ maps of Fig. \ref{fig:3}a are derived once more with respect to $j$, resulting in the $\dVdIdx$ maps of Fig. \ref{fig:Scompare}. In these maps along the full length of the device, the hydraulic jump is the dip-hump feature indicated by the white dashed box in every panel, and in more detail in Fig. \ref{fig:Szoom}. The dip-hump feature clearly moves to different current densities when the gate bias is changed, is only present on one side of the device, and moves to the other side when the source and drain are swapped (lower left panel). Also, the feature disappears when a magnetic field is applied (lower right panel). Introducing a magnetic field causes a Lorentz force on the carriers, breaking momentum conservation and qualitatively modifying the resulting current flows.
\clearpage

\subsection{Simultaneous local and global transport probes}
\label{simul}

It is possible to derive $V_{KPFM}(x,I)$ maps with respect to $I$. The resulting $\frac{dV}{dI}(x,j)$ can then be interpreted as a \textit{local} $\frac{dV}{dI}(j)$ measurement, for many locations along the device. These calculated $dV/dI$ are shown in Fig. \ref{fig:S6} as the many traces from orange to purple (corresponding to locations indicated on the image of the device). The $dV/dI$ traces from simultaneous, global transport at contacts 1 and 2 are shown as the green ($dV_1/dI$) and black ($dV_2/dI$) traces.

The transport feature, the maximum in $dV/dI$, linked to the hydraulic jump is indicated by the orange arrow in the top right. It is only present for specific locations on the left of the device, and then suddenly disappears at the hydraulic jump (orange arrow on the device). Similarly, the hydraulic jump only is visible in transport at contact 1 (green trace), and not present at contact 2 (black trace). This confirms that the global electronic transport feature can be located to a single location inside of the device with KPFM.

\begin{figure}
    \centering
    \includegraphics[width=0.95\linewidth]{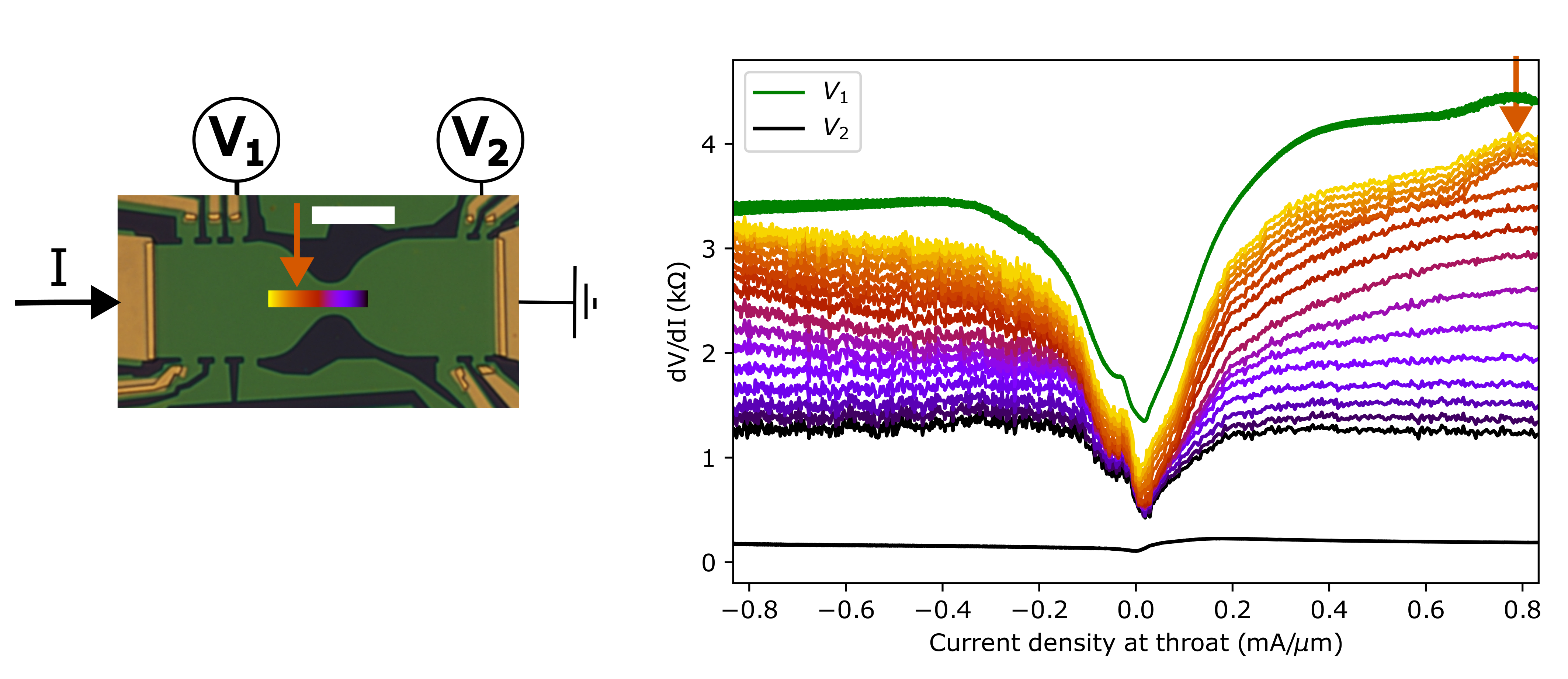}
    \caption{
    \textbf{Locating electronic features in local and global transport}
    Local $dV/dI (j)$ measurements along the device (yellow to purple colorbar) with simultaneous transport (contact 1 in green, contact 2 in black). The location where the transport feature (orange arrow) appears is indicated on the device (orange arrow).}
    \label{fig:S6}
\end{figure}
\clearpage

\clearpage
\subsection{Removing work function features in KPFM}
\label{SI:work}
In this section, we demonstrate the presence of both work function and electrical potential features in the KPFM data. At $j=0$, the KPFM features arise from work function variations on the order of 100~mV (Fig.~\ref{fig:Deltav}a, d). At high bias, the potential drop from the source bias dominates the data (Fig.~\ref{fig:Deltav}c, e). Because we do not expect work function variations to change with $j$, subtracting data collected for different $j$ removes work function contributions, which is critical for analyzing spatial derivatives such as $\dVdx$.

\begin{figure}[h!]
    \centering
    \label{fig:Deltav}
    \includegraphics[width=0.9\linewidth]{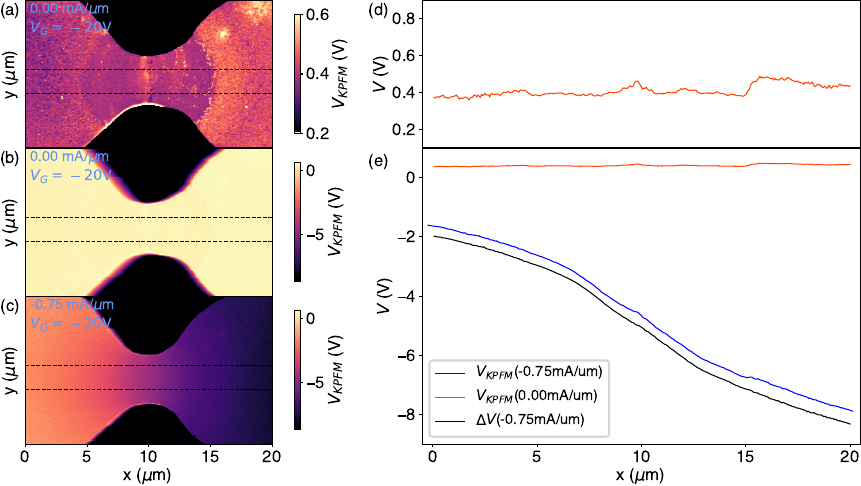}
    \caption{
    \textbf{Calculating $\Delta V$.} 
    \textbf{a-c.} KPFM images with current and gate voltages indicated on each panel. The values outside of the bilayer graphene nozzle exceed the colour scale limits as a result of the gate. The same data is shown in a and b, but with a smaller color scale in a to reveal work function features. The data in panel c is repeated from Fig.~\ref{fig:2}b of the main text but plotted with a slightly different color scale (to accommodate the data in b). The color scale is identical between b and c for comparison. The dashed lines indicate the range of data used to calculate the profiles in d-e.
    \textbf{d-e.} KPFM profiles, $\Vkpfm(j)$, taken from images in a-c (blue and orange). The difference $\Delta V(j) = \Vkpfm(j) - \Vkpfm(0)$, used to remove work function features (black), is repeated from Fig.~\ref{fig:2} of the main text and plotted on the same axes for comparison. The profiles were calculated by taking the median of the scan lines through the center of the nozzle indicated by the dashed lines in a-c. Median rather than mean was used to reduce the influence of variations in the y direction. Profiles have been aligned in $x$ based on features in the simultaneously acquired topographies. The $j=0$ profile is repeated in d and e, but on a smaller vertical scale in d to reveal the work function features.
    }
\end{figure}

\begin{comment}
\newpage
\subsection{Locally tracing $dV/dI$ to the flattened $V(x)$}

\begin{figure}
    \centering
    \includegraphics[width=0.9\linewidth]{image/FigS6v1.png}
    \caption{\textbf{Locally resolved dV/dI (x)}. Traces for dV$_{KPFM}$/dI at positions along the nozzle (colourbar from yellow to black in the device image) are compared with dV/dI traces measured via transport (green and black traces).}
    \label{fig:S6}
\end{figure}
\end{comment}

\clearpage
\subsection{Navier-Stokes Solution Methods}

%To confirm that the flattening of the electrochemical potential $V(x)$ corresponds to a hydraulic jump, we performed extensive numerical simulations of the compressible electronic flow with a minimal hydrodynamic model, assuming a single circular Fermi surface with parabolic dispersion \cite{lucas2018}.  This model is appropriate for bilayer graphene so long as the temperature remains small compared to the Fermi temperature throughout the device.
%We solved Eqs.~\eqref{eq:ns} numerically until steady-state was reached using finite volume methods . % \andy{how were simulation parameters chosen?} 
%The quasistatic treatment of $\phi$ in this model makes flow parameter discontinuities at the shock only appear in the \textit{electrochemical} potential {$V(x) = (\phi(x) - \mu(n(x))/e$} through density $n$, resulting in the consistent underestimation of the magnitude of the shock. A more realistic model will need to incorporate nonlocal Coulomb coupling between $n(x)$ and $\phi(x)$, as well as the contribution of thermal gradients through the Seebeck effect.

In this section, we explain the numerical methods used in the one-dimensional and two-dimensional toy models for the Fermi liquid regime of a bilayer-graphene nozzle at $T=0$.  In both one and two dimensions, our methods essentially solve the Navier-Stokes equation with external drive, momentum relaxation, and the pressure function for bilayer graphene at $T=0$,
\begin{align}
    P(n) = \frac{\pi \hbar^2}{4m^2}n^2.
\end{align}
For the simulations, we operate at $\mu > 0$ so that only $n$-type carriers realize.  Current flows in the $+\vu{x}$ direction, with electrons correspondingly moving in the $-\vu{x}$ direction.

When currents are small (linear response), \eqref{eq:ns} form a closed system since $\grad \phi$ and $n$ do not need to be independently determined ($\nabla P$ and $\nabla n$ are proportional). In the nonlinear regime, when the convective term $n \vb{v}_d\cdot\grad\vb{v}_d$ is large, the local electric potential $\phi(\vb x)$ must be specified independently of $n$ (or self-consistently though Maxwell's equations).  Since the usual gradual-channel relation $en(\vb x) = C \cdot \phi(\vb x)$ does not hold for a lightly doped, frozen out silicon backgate, we adopt a mean-field treatment of the potential $\phi(\vb x)$, fitting an Ohmic potential profile to the KPFM measurements and using the resulting $\phi(\vb x)$ as a static driving field for \eqref{eq:ns}.
Accordingly, the electronic speed of sound is effectively set by the electronic contribution (in the absence of the gate), consistent with the experimental data in Figure \ref{fig:2}.

\label{numerical}
\subsubsection{One-dimensional model}\label{sec:1dmodel}
To solve \eqref{eq:1dns}, we partitioned the time evolution of the first order PDE for the two hydrodynamic variables $q = (n, n v_d)$
according to 
\begin{align}
    \pd_t \begin{pmatrix}
        n\\
        n v_d
    \end{pmatrix} + \pd_x F  = S_{\mathrm{geom}}+ S_{\mathrm{mr}} + S_{\mathrm{visc}}.
\end{align}
The left-hand-side is a 1D conservation law with ``numerical flux function''~\cite{LeVeque2002} $F$ given by 
\begin{align}
    F = \begin{pmatrix}
        n v_d \\
        \frac{(n v_d)^2}{n} + \frac{\pi \hbar^2}{4m^2}n^2
    \end{pmatrix},
\end{align}
related to the 1D shallow water equations~\cite{LeVeque2002}. 

The right hand side features three contributions, namely nozzle geometry,
\begin{align}
    S_{\mathrm{geom}} = \begin{pmatrix}
        - n v_d \frac{\pd_x W}{W}\\
        -\frac{(n v_d)^2}{n} \frac{\pd_x{W}}{W}
    \end{pmatrix};
\end{align}
momentum relaxation (and external forcing),
\begin{align}
    S_{\mathrm{mr}} = \begin{pmatrix}
        0 \\
        \frac{1}{m}n e \pd_x \phi -\gammamr(n) n v_d
    \end{pmatrix};
\end{align}
and viscosity,
\begin{align}
    S_{\mathrm{visc}} = \begin{pmatrix}
        0\\
        \frac{1}{m} \frac{1}{W}\pd_x (\eta W \pd_x v_d)
    \end{pmatrix}.
\end{align}

The experimentally measured differential resistance $\dv*{V}{I}$ does not depend on carrier density $n_0$ (Fig. \ref{fig:1}b) at large current density. We phenomenologically capture this behaviour, assuming the resistance is dominated by the Ohmic background, by introducing a density-dependent momentum-relaxation rate $\gammamr (n) = \alpha n$, with $\alpha$ constant and determined from the measurement, resulting in a density-independent Drude resistivity.

We solved the left-hand-side---the conservative part---using finite volume methods to capture shock discontinuities in the limit $\eta \rightarrow 0$.  In particular, we used Roe's approximate Riemann solver~\cite{LeVeque2002} with minmod slope limiting for a high resolution method, as in \cite{mendl2021b, farrell2022}.  We incorporated the source terms with nested Strang splitting~\cite{LeVeque2002} with a fourth order Runge-Kutta scheme for the time-stepping. We fixed a reference density $n_0 \sim 10^{12}$ cm$^{-2}$ and reference length $L = 40 \mu$m corresponding to the device size.  We picked a reference velocity $v_0 = v_s(n_0) = \hbar/m \sqrt{\pi n_0 / 2}$.  In units where $n_0 = v_0 = L = 1$, our simulations used time step $\dd{t} = 1 / 10000$ and $\dd{x} = 1 / 500$.  The small value of $\dd{x}$ was necessary to capture the $\sim 200$ nm feature observed with KPFM. The small $\dd{t}$ was required to resolve the relatively large momentum relaxation rate, $\gamma_{mr}(n_0) \sim 700$ in units of $v_0 / L$. Compressible flow nonetheless persists due to $\gammamr(n) = \alpha n$ for constant $\alpha$. Fig.~\ref{fig:S8} shows example traces of $\frac{d}{dx}\frac{\Delta V}{\Delta I}$ as $\gamma_{mr}$ increases.

\begin{figure}
    \centering
    \includegraphics[width=0.95\linewidth]{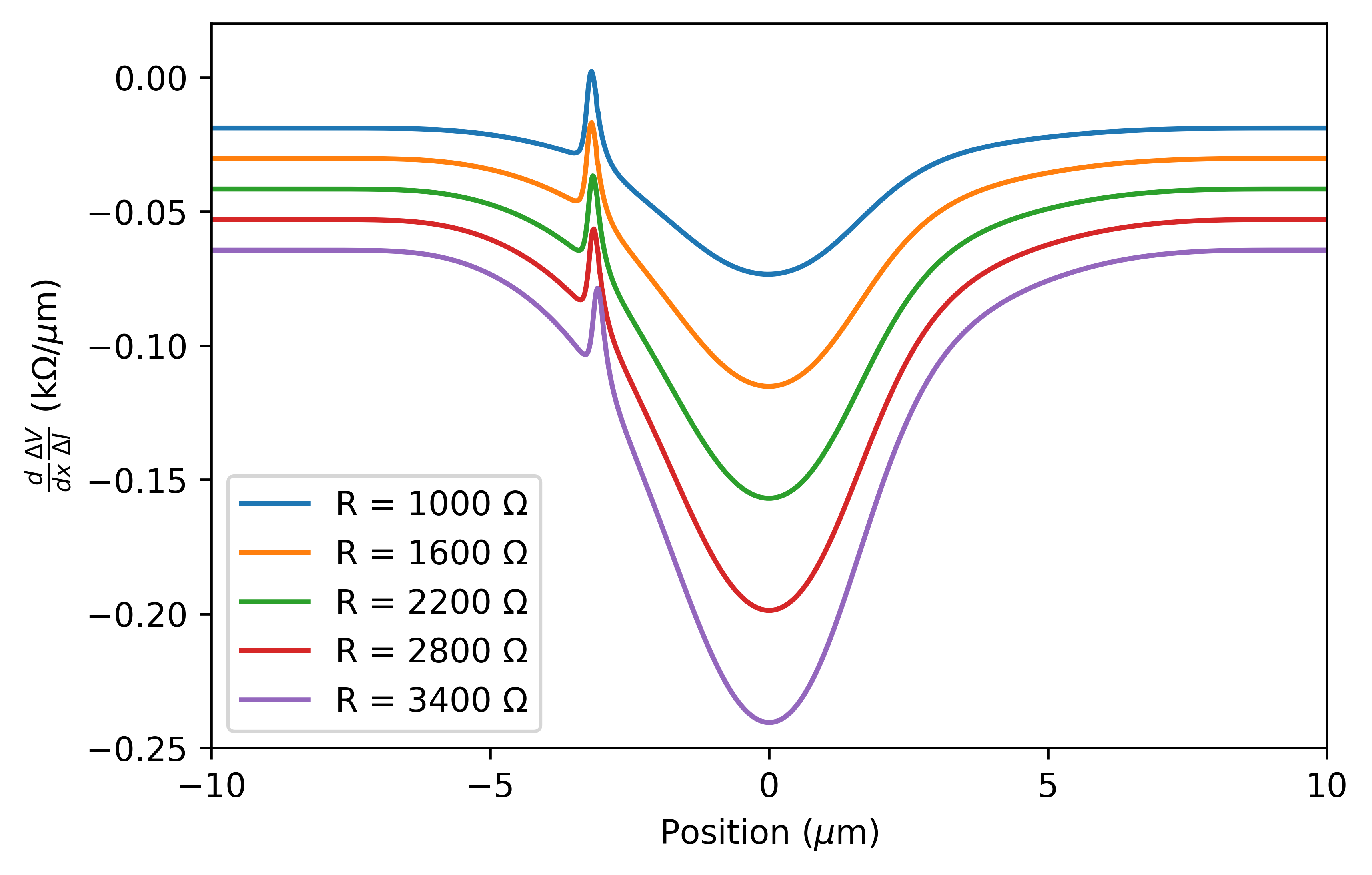}
    \caption{\textbf{Effect of momentum relaxation on shock magnitude}
    Calculated local differential resistivity for different values of Drude resistance $R$, the degree of momentum relaxation.}
    \label{fig:S8}
\end{figure}

\subsubsection{Two-Dimensional Model}
We similarly solved \eqref{eq:ns} using a uniform grid and the same nondimensionalization as the 1D case. In 2D, we estimated the electric potential $\phi$ driving the flow by solving $\grad^2 \phi = 0$ with the normal gradient $\grad \phi \cdot \vu{n} = 0$.  The two dimensional shallow-water equations were solved using dimensional splitting~\cite{LeVeque2002}.  In this case, the hydrodynamic variables are $n, n v_x, n v_y$, where we have dropped the subscript $d$ on $\vb v_{d}$ to avoid cluttering the notation.  We partitioned the time-evolution of the hydrodynamic variables according to
\begin{align}\label{eq:2dmethod}
    \pd_t \mqty( n\\nv_x \\nv_y) + \pd_x F + \pd_y G = S_{\mathrm{mr}} + S_{\mathrm{visc}}.
\end{align}
The left-hand side is a 2D conservation law with numerical flux functions~\cite{LeVeque2002} $F$ and $G$ given by:
\begin{align}
    F = \mqty(nv_x \\ \frac{(nv_x)^2}{n} + \frac{\pi \hbar^2}{4m^2}n^2 \\ \frac{(n v_x) (n v_y)}{n}),\ G = \mqty(n v_y \\ \frac{(nv_x)(n v_y)}{n} \\ \frac{(n v_y)^2}{n} + \frac{\pi \hbar^2}{4m^2}n^2).
\end{align}
We used Strang splitting~\cite{LeVeque2002} on the conservation law to form a simple dimensional-splitting method, which sufficed for our purposes to observe the two-dimensional electronic shock.

The source terms on the right hand side of \eqref{eq:2dmethod} respectively come from momentum relaxation (and external drive), 
\begin{align}
    S_{\mathrm{mr}} = \mqty(0 \\ \frac{1}{m}ne\pd_x \phi - \gamma(n) n v_x \\ \frac{1}{m}ne \pd_y \phi - \gamma(n) n v_y);
\end{align}
and viscosity, 
\begin{align}
    S_{\mathrm{visc}} = \mqty(0 \\ \frac{\eta}{m} \grad^2 v_x \\ \frac{\eta}{m} \grad^2 v_y);
\end{align}
where $\grad^2 \equiv \pd_x^2 + \pd_y^2$ is the Laplacian in 2D.  We used nested Strang splitting with fourth-order Runge-Kutta time-stepping to incorporate the source terms.  For the full 2D simulation, we used $\gammamr =$constant, calibrated to the $I=0$ mA measured resistance to visualize the shape of the shock front, which explains scale difference.  Otherwise, the two-dimensional simulations used the same parameters as the one-dimensional reduction described in Sec.~\ref{sec:1dmodel}.

These subtleties related to boundary conditions and electric field suggest that future quantitative models of compressible flow in 2DEGs could benefit from direct solution of the 3D electrostatics in the gate-oxide-graphene complex, beyond the scope of the present paper.

\subsubsection{Boundary conditions}
In this section we explain our choice of boundary conditions for the simulation results from the main text.  It is worth stating directly that the boundary conditions for electronic hydrodynamics are subtle in general~\cite{lucas2018}---while the simple scheme we have picked is self-consistent and numerically stable, more sophisticated choices may help quantitative agreement with future compressible flow experiments.

Assuming both inlet and outlet are subsonic, the hyperbolic one-dimensional PDE requires one boundary condition on each of the left hand side (outlet) and right hand side (inlet) of the simulation domain.  Since the electrochemical potential $\mu_{ec}$ is fixed at the inlet in the experiment for $n$-type carriers, we fixed $n = \mathrm{constant}$ at the inlet (right hand side). At the outlet (left hand side), we impose ``do-nothing'' (non-reflective) boundary conditions, allowing density $n$ to float at the outlet while setting the incoming and outgoing characteristics equal.  Since the experiment is current-driven, we adjust the magnitude of the driving field $\pd_x \phi$ to reach the desired current.

In two dimensions, we require one additional boundary condition at both the inlet and the outlet, which are supplied by demanding $v_y = 0$ on either end of the channel.  Finally, on the, the nozzle walls (vacuum), viscous flow motivates a no-slip condition: $\vb{v} = 0$ as a vector on the wall.

\putbib[nozzle-bibtex]
\end{bibunit}

\end{widetext}

\end{document}